\pdfoutput=1
\documentclass[numberedappendix, fleqn, twocolumn, appendixfloats, twocolappendix]{openjournal}

\usepackage{amsmath}
\usepackage{graphicx}
\usepackage{booktabs}
\usepackage{threeparttable}
\usepackage{needspace}

\usepackage{hyperref}
\hypersetup{
  colorlinks=true, 
  linkcolor=blue, 
  citecolor=blue,  
  filecolor=blue, 
  urlcolor=blue}

\newcommand{\fTRGB}{f_{\mathrm{T}}}
\newcommand{\LTRGB}{L_{\mathrm{T}}}
\newcommand{\mTRGB}{m_{\mathrm{T}}}
\newcommand{\Nexpected}{N_0}
\newcommand{\fmin}{f_{\mathrm{min}}}
\newcommand{\fmax}{f_{\mathrm{max}}}
\newcommand{\Nbar}{\bar{N}}
\newcommand{\Nabove}{N_+}
\newcommand{\Nbelow}{N_-}
\newcommand{\rhobelow}{\rho_{-}}
\newcommand{\rhoabove}{\rho_{+}}
\newcommand{\sigmaftrgb}{\sigma_{\mathrm{T}}}
\newcommand{\rhostar}{\rho_{*}}
\newcommand{\fThat}{\hat{f}_{\mathrm{T}}}
\newcommand{\fcut}{f_{\mathrm{cut}}}
\newcommand{\microjansky}{\textmu Jy}
\newcommand{\mgalaxy}{\bar{m}}
\newcommand{\mjfield}{m_{j}}
\newcommand{\msetjfield}{m_{1:j}}
\newcommand{\Dsetj}{D_{1:j}}
\newcommand{\MTRGBHSTband}{{M}_{\mathrm{F814W}}^{\mathrm{TRGB}}}
\newcommand{\neff}{n_{\mathrm{eff}}}
\newcommand{\fTtrue}{\tilde{f}_{\mathrm{T}}}
\newcommand{\rhobelowtrue}{\tilde{\rho}_{-}}
\newcommand{\rhoabovetrue}{\tilde{\rho}_{+}}
\newcommand{\ch}[1]{\multicolumn{1}{c}{#1}}

\begin{document}

\title{Bayesian Inference of the Tip of the Red Giant Branch: \\
I. One-Dimensional Analysis
  \vspace*{-15mm}}

\shorttitle{Bayesian Inference of the TRGB}

\author{$\!$James D.\ Gibbon$^{1\ast}$ and
Daniel J.\ Mortlock$^{1,2}$}

\affiliation{$^1$Department of Physics, Imperial College London, London SW7 2AZ, United Kingdom \\
$^{2}$Department of Mathematics, Imperial College London, London SW7 2AZ, United Kingdom}
\email[$^\ast$]{james.gibbon25@imperial.ac.uk}
\shortauthors{Gibbon \& Mortlock}

\begin{abstract}
\noindent
The tip of the red giant branch (TRGB) is a relative distance measure that can be used, as part of a cosmic distance ladder, to infer the Hubble constant. We present a Bayesian framework for TRGB inference that incorporates selection effects and propagates per-star flux uncertainties, modelling the observed catalogue as a realisation of an inhomogeneous Poisson point process which includes the RGB and contamination from the asymptotic giant branch (AGB). We derive expressions for the predicted TRGB uncertainty in terms of the photometric noise, stellar density fainter than the tip, and the ratio of the AGB to AGB+RGB number densities, \(r\). We validate these results using synthetic data from a simplified luminosity function. Applying the full model to synthetic catalogues, we find that the dependence on \(r\) is the strongest of the three, and that reliable inference requires \(r\lesssim0.5\); otherwise, the posterior can be multimodal, and the TRGB cannot reliably be identified. We then fit data from 11 \textit{Hubble Space Telescope} fields of the calibrator galaxy NGC~4258. We find intrinsic field-to-field scatter of \(\tau=0.067^{+0.023}_{-0.017}\)~mag (beyond the measurement noise), and a galaxy-level, extinction-corrected TRGB magnitude of \(25.324 \pm 0.026\) in the F814W filter. Combining this with the maser distance modulus to NGC~4258 of \(29.397 \pm 0.032\)~mag, we determine the absolute magnitude of the TRGB to be \({M}_{\mathrm{F814W}}^{\mathrm{TRGB}} = -4.073 \pm 0.026\;\text{(tip)} \pm 0.032\;\text{(dist)}\), consistent with, but brighter than, other recent analyses of the same data. 
\end{abstract}


\section{Introduction}

Recent high-precision measurements of the Hubble constant, \(H_0\), which parametrises the present-day expansion of the Universe, have revealed a persistent and significant discrepancy between early- and late-universe probes, commonly referred to as ``the Hubble tension''. Early-universe constraints, primarily from measurements of the cosmic microwave background combined with the assumption of a standard flat Lambda cold dark matter (\(\Lambda\)CDM) cosmological model, yield values such as \(H_0 = 67.4 \pm 0.5 \, \mathrm{km\,s^{-1}\, Mpc^{-1}}\) \citep{collaboration_planck_2020}. In contrast, late-universe distance-ladder measurements based on Cepheid-calibrated Type Ia supernovae yield higher values, including \(H_0 = 73.04 \pm 1.04 \, \mathrm{km\,s^{-1}\, Mpc^{-1}}\) \citep{riess_comprehensive_2022}. Alternative distance-ladder calibrations, including those based on the tip of the red giant branch (TRGB), do not resolve this discrepancy: analyses anchored to NGC~4258 yield \(H_0 = 70.39 \pm 1.22 \, (\mathrm{stat}) \,\pm 1.33 \, (\mathrm{sys}) \pm 0.70 \, (\sigma_{\mathrm{SN}})\) \citep{freedman_status_2025}, \(71.5\pm1.8\) \citep{anand_comparing_2022}, and \(73.22\pm2.06\,\mathrm{km\,s^{-1}\, Mpc^{-1}}\) \citep{scolnic_cats_2023}. The disagreement between early- and late-universe values implies either unaccounted-for systematic uncertainties or a need for extensions to the standard cosmological model. 

Tensions between independent calibrations of the distance ladder's intermediate rungs motivate the development of unbiased distance indicators with well-characterised uncertainties. Whilst the TRGB has been widely used for distance estimation \citep{da_costa_standard_1990, lee_tip_1993, madore_tip_1995, salaris_tip_1997, rizzi_tip_2007, jang_tip_2017}, and has been further refined in modern distance-ladder analyses \citep{wu_comparative_2023, hoyt_chicago_2026}, most implementations infer the TRGB magnitude using edge-detection applied to binned star counts, with or without smoothing. These approaches typically compress the photometric data into a single histogram, thereby discarding information about individual stellar measurements and their uncertainties. 

Existing likelihood-based methods \citep{mendez_deviations_2002, makarov_tip_2006, conn_bayesian_2011} fit a parametric luminosity function (LF) rather than searching for an edge in a smoothed histogram, and \cite{conn_bayesian_2011} place this in a Bayesian framework. These are concrete steps towards a fully Bayesian treatment, but two approximations remain, which limit what can be inferred. First, these works model the photometric scatter with a population-level noise distribution evaluated at each magnitude rather than using per-star uncertainties. This treats every star at a given magnitude as equally precise, so an unusually noisy measurement carries the same weight as a clean one. Second, the LF is normalised over the fitted range, so the number of stars in the catalogue is conditioned upon rather than modelled, discarding the information it carries about the stellar density.

Here we develop a fully Bayesian approach to TRGB inference, explicitly modelling observational selection and per-star measurement uncertainty. In this paper, we consider only the one-dimensional case of fluxes/magnitudes along the RGB; we will subsequently extend this to the two-dimensional case in which colour information is included as well. Section~\ref{sec:TRGB} introduces the stellar physics of the TRGB and derives the LF used to infer the TRGB. Section~\ref{sec:statistical model} develops the statistical framework underlying TRGB inference, outlining the generative model of the synthetic dataset and the Poisson process that we use to model the TRGB. We forecast the error in the TRGB luminosity using a simplified LF in Section~\ref{sec:error forecast}. Section~\ref{sec:sampling} describes the posterior sampling process, and the model is tested with synthetic data in Section~\ref{sec:coverage}. Section~\ref{sec:Heteroscedastic Photometric Uncertainty} introduces a heteroscedastic model of the photometric uncertainty, in line with observations, before the model is applied to real data from NGC~4258 in Section~\ref{sec:Real data}. We discuss the findings and limitations of the model in Section~\ref{sec:Discussion}, before concluding in Section~\ref{sec:conclusion}.

\section{The Tip of the Red Giant Branch}\label{sec:TRGB}

Observationally, the TRGB is seen as a sharp discontinuity in the number of low-mass (\(M \lesssim 2\,M_\odot\)) first-ascent red giant stars \citep{salaris_red_2002}, marking the onset of helium burning in their cores. These stars fuse hydrogen in a shell surrounding an electron-degenerate helium core, ascending the RGB over \(\sim\!100\)~Myr \citep{mocak_core_2008, hidalgo_updated_2018}. When the mass of the He core reaches \(0.5\,M_{\odot}\), the density and temperature in regions of the core are high enough to support He fusion via the triple-alpha process \citep{salpeter_nuclear_1952}, causing a thermal runaway reaction, the ``He flash''. The He flash event marks the end of the star's RGB phase, initiating stable He fusion in the core. The star contracts, reduces in luminosity, and heats, moving downwards and leftwards in the colour--magnitude diagram (CMD), onto the horizontal branch over \(\sim\!0.5\)~Myr \citep{hidalgo_updated_2018}. The timescales involved produce a sharp truncation of the RGB in the CMD. In the \(I\) band --- and its \textit{Hubble Space Telescope} (\textit{HST}) ACS equivalent F814W, used throughout this work --- the TRGB magnitude has been found to be insensitive to metallicity for metal-poor populations (\([\mathrm{Fe/H}] < -0.7\)), whilst the bolometric luminosity of the TRGB varies by only \(\sim\!0.1\)~mag over ages of 2--15~Gyr at fixed metallicity, making it a standardisable candle \citep{iben_asymptotic_1983, lee_tip_1993}.

To undertake Bayesian inference of the TRGB, we require models for both the RGB (Section~\ref{sec:RGB_LF}) and the asymptotic giant branch (AGB) contaminant population (Section~\ref{sec:AGB_LF}), which are then combined into a joint LF (Section~\ref{sec:Full LF}). 

\subsection{The red giant branch luminosity function}
\label{sec:RGB_LF}

Stars in the RGB satisfy the core mass--luminosity (CML) relation, \(L\propto M_\mathrm{c}^\gamma\) \citep{eggleton_evolution_1968, castellani_luminosity_1989, salaris_red_2002}. The core mass, \(M_\mathrm{c}\), rises as \(\dot M_\mathrm{c} \propto L\) as fusion in the shell deposits He onto the core, which in turn increases the luminosity according to the CML relation as \(\mathrm{d}L/\mathrm{d}M_\mathrm{c} \propto M_\mathrm{c}^{\gamma-1}\).

Combining these relationships implies
\begin{equation}
   \dot L = \frac{\mathrm{d}L}{\mathrm{d}M_\mathrm{c}}\dot M_\mathrm{c} \propto M_\mathrm{c}^{\gamma-1} L \propto L^{(\gamma-1)/\gamma} L = L^p,
\end{equation}
with \(p = 2 - 1/\gamma >1\), implying a runaway process as stars approach the tip luminosity, \(\LTRGB\).

Provided the stars enter the RGB at a constant rate, the number of stars at a given luminosity is proportional to the time a star spends with that luminosity, so 
\begin{equation}
\label{eqn:dndl}
\frac{\mathrm{d}N}{\mathrm{d}L} \propto \frac{\mathrm{d}t}{\mathrm{d}L} \propto \Theta(\LTRGB-L)\,L^{-p},
\end{equation}
where the Heaviside step function \(\Theta(\cdot)\) encodes the fact that \(L \leq \LTRGB\). The normalisation is not set by this argument, as it depends on the number of RGB stars in the observational area and enters the model as the parameter \(\rhobelow\) in Equation~\eqref{eqn:full LF}.

For stars at the same distance in a single host galaxy, Equation~\eqref{eqn:dndl} translates directly to fluxes as 
\begin{equation}\frac{\mathrm{d}N}{\mathrm{d}f} \propto \Theta(\fTRGB-f) \, f^{-p}.
\end{equation}
In terms of magnitudes, defined in terms of a reference flux \(f_0\) by \(m = -5/2 \, \log_{10}(f/f_0)\), this becomes
\begin{equation}\frac{\mathrm{d}N}{\mathrm{d}m} \propto \Theta(m-\mTRGB)\, 10^{q(m-\mTRGB)},
\end{equation}
where \(q \equiv 2/5 \,(p-1)\). 

The above derivation is equivalent to that presented by \cite{castellani_luminosity_1989}, who obtain the same result working directly with magnitudes and predict \(q=0.32\pm0.02\). The same power-law form has been used in subsequent TRGB studies, where it is adopted empirically rather than derived: \cite{mendez_deviations_2002} measure \(q=0.30\pm0.04\) from their data and fix this value; and \cite{makarov_tip_2006} assume the exponential form but leave the exponent free.

Using stellar evolutionary tracks from the Bag of Stellar Tracks and Isochrones (BaSTI) database \citep{hidalgo_updated_2018}, we verify that the theoretical RGB luminosity evolution follows a power law \(\dot{L} \propto L^p\) for \(L \leq \LTRGB\). The relation is well described by a single exponent \(p \simeq 1.67\), which gives \(q \simeq 0.27\). The corresponding CML exponent is \(\gamma\simeq3\), which is lower than the expected \(\gamma\simeq 7\) \citep{boothroyd_low-mass_1988}. However, the observed counts constrain \(\gamma\) only weakly: \(\gamma\simeq 7\) corresponds to \(q\simeq0.34\), and both this value and our \(q\simeq0.27\) lie within \(\sim\!1\sigma\) of the slope measured by \cite{mendez_deviations_2002}. More importantly, the value of \(\gamma\) does not affect the inference as the LF slope enters the model as a free parameter rather than a fixed prediction.

\subsection{The asymptotic giant branch luminosity function}
\label{sec:AGB_LF}

Inconveniently, the position of the TRGB in colour--magnitude space is also populated by AGB stars, which act as a contaminant population that makes it more difficult to measure the TRGB. There is no equivalent physical derivation for the AGB demographics, so we instead follow \cite{mendez_deviations_2002} and \cite{makarov_tip_2006} in asserting the same functional form for the AGB counts, leaving its slope as a free parameter. Our one-dimensional treatment of the problem here does not account for the fact that the AGB also lies at an angle to the RGB in the CMD, making it an intrinsically two-dimensional object. 

The relative number of AGB and RGB stars at the tip empirically depends on the position within a galaxy, which is discussed further as a survey-design consideration in Section~\ref{sec:Survey}.

\subsection{The joint luminosity function}
\label{sec:Full LF}

The natural joint model for stars in the vicinity of the TRGB would be a sum of the RGB and AGB populations. However, the lack of a compelling functional form for the AGB means that a reasonable practical option is to use an LF\footnote{Although the population model used here is defined in terms of fluxes rather than luminosities, we refer to it as an LF. This could be made more formally correct by working with luminosities and including the distance to the galaxy, but this would be degenerate with the TRGB luminosity.} defined by
\begin{equation}
    \psi(f,\theta) =
\begin{cases}
\rhobelow\left(\frac{f}{\fTRGB}\right)^{-a} & \fmin \leq f\leq\fTRGB\\
\rhoabove\left(\frac{f}{\fTRGB}\right)^{-b} & f>\fTRGB,
\end{cases}
\label{eqn:full LF}
\end{equation}
where \(\theta = (\fTRGB, a, b, \rhobelow, \rhoabove)\) is the list of population-level parameters, and \(\rhobelow\) and \(\rhoabove\) are the stellar densities immediately below (RGB+AGB) and above (AGB only) the tip. The lower bound at \(\fmin\) is required to ensure that any integrals calculated are finite. Provided \(\fmin\ll\fTRGB\), the inference of \(\fTRGB\) should be insensitive to its adopted value. Equation~\eqref{eqn:full LF} is equivalent to the model used by \cite{mendez_deviations_2002} and \cite{makarov_tip_2006}, but expressed in terms of fluxes rather than magnitudes.

A key quantity that determines how well the tip can be measured is the strength of the ``break'' in the stellar counts at the TRGB. This is characterised here by the ratio of the AGB to the AGB+RGB densities at \(f = \fTRGB\), 
\begin{equation}
    r \equiv \frac{\rhoabove}{\rhobelow},
\label{eqn:r definition}
\end{equation}
which can range between 0, corresponding to the ideal situation of no AGB at all, and 1, in which there are no RGB stars and hence no feature to measure.

\section{Statistical Model}
\label{sec:statistical model}

We adopt a Bayesian approach, inferring all unknown quantities simultaneously, before marginalising over the joint posterior to obtain constraints on the parameters of interest that properly propagate every source of uncertainty. The ultimate goal of this work is to infer the posterior distributions of the population-level parameters, \(\theta\). This requires jointly inferring the latent true fluxes, \(f_{1:N}\), of the \(N\) stars in the catalogue. These are related to their corresponding measured flux via a generative measurement model (Section~\ref{sec:Generative Model}), from which we obtain the resultant posterior distribution (Section~\ref{sec:posterior distribution}).

\subsection{Generative model}
\label{sec:Generative Model}
Following \cite{streit_poisson_2010} and \cite{mandel_extracting_2019}, we treat the observed stellar catalogue as a realisation of an inhomogeneous Poisson point process (PPP). The Poisson distribution is not an additional assumption here: if the stars are scattered independently in flux, with finite counts in every flux region, the PPP emerges directly \citep[Section~2.9.3]{streit_poisson_2010}. 

We start by describing a generative model to sample from the LF, \(\psi(f, \theta)\). First, we draw the number of stars produced by the chosen LF from the Poisson distribution,
\begin{equation}
    \Nexpected \sim \mathrm{Poisson}(\Nbar_{\mathrm{LF}}),
\end{equation}
where
\begin{equation}
    \Nbar_{\mathrm{LF}} \equiv \int_{0}^{\infty} \mathrm{d} f \, \psi(f, \theta)
    \label{eqn:N_LF}
\end{equation}
gives the expected number of stars in the population before the measurement model and selection are applied (and which must hence be finite). For each star, indexed by \(i \in \{1, 2, \ldots, \Nexpected\}\), we draw a true flux from the normalised LF,
\begin{equation}
    f_i \sim \frac{\psi(f, \theta)}{\Nbar_{\mathrm{LF}}},
    \label{eqn:true_fluxes}
\end{equation}
and apply the measurement model
\begin{equation}
    \hat{f}_i \sim P(\hat{f} \mid f_i),
\end{equation}
which defines the sampling distribution of the measured flux for star \(i\). The specific form of \(P(\hat{f} \mid f)\) depends on the noise model; this could be a normal distribution with constant-in-flux variance \(\sigma^2\), a normal distribution with flux-dependent variance \(\sigma^2(f) = \sigma_0^2 + Cf\) (explored more fully in Section~\ref{sec:Heteroscedastic Photometric Uncertainty}), or could incorporate other stochastic effects such as Poisson photon counts.

Finally, a flux cut at \(\fcut\) is applied, and the \(N\) stars with \(\hat{f} \geq \fcut\) enter the final catalogue, \(D = (\hat{f}_1, \hat{f}_2, \ldots, \hat{f}_N)\). For constant-in-flux noise, \(\fcut=\rhostar\sigma\), where \(\rhostar\) is the signal-to-noise ratio (SNR) of the cut. For flux-dependent noise, \(\fcut\) remains a constant threshold, but its value is determined by the form of the noise.

Whilst the above algorithm is intuitive, in that it mimics the actual data-generation process, it is not immediately clear how to obtain the resultant posterior distribution from it. We can refactor the problem to lead us to the final posterior distribution by reducing the multi-step data-generation and selection process to a single equivalent PPP. In the generative model, each of the \(\Nexpected\) stars independently survives the selection cut with probability \(P(S \mid f_i, \fcut)\). The number of survivors is therefore a sum of independent Bernoulli trials applied to a Poisson-distributed number of stars. Thinning the PPP in this way results in another PPP, and hence the number of selected stars is drawn from a Poisson distribution,
\begin{equation}
    N \sim \mathrm{Poisson}(\Nbar),
    \label{eqn:N draw}
\end{equation}
where
\begin{equation}
    \Nbar(\theta, \fcut) = \int_0^\infty \mathrm{d} f \, \psi(f, \theta)\, P(S \mid f, \fcut),
    \label{eqn:N_bar}
\end{equation}
and
\begin{equation}
    P(S \mid f, \fcut) = \int_{\fcut}^{\infty} \mathrm{d} \hat{f}\, P(\hat{f} \mid f)
    \label{eqn:P(S|f)}
\end{equation}
is the probability that a star with true flux \(f\) is selected into the catalogue. 

The pre-selection PPP intensity on \((f, \hat{f})\) is \(\psi(f, \theta)\, P(\hat{f} \mid f)\). Conditioning on the selection \(\hat{f} \geq \fcut\) introduces the Heaviside step function \(\Theta(\hat{f} - \fcut)\), and normalising over all \((f, \hat{f})\) gives the denominator \(\Nbar\). This leaves the joint distribution of the true and measured flux of each selected star as
\begin{equation}
    P(f, \hat{f} \mid \theta, \fcut, S) = \frac{\psi(f, \theta)\, P(\hat{f} \mid f)\, \Theta(\hat{f} - \fcut)}{\Nbar(\theta, \fcut)}.
    \label{eqn:joint_draw}
\end{equation}

\needspace{4\baselineskip}
\subsection{Posterior distribution}
\label{sec:posterior distribution}
Our inference target is the joint distribution on the population-level parameters and the latent true fluxes, given a catalogue of \(N\) observed fluxes, their uncertainties, and a selection function. Using Bayes' theorem, we can write down the posterior, up to normalisation, as 
\begin{align}
    &P(\theta, f_{1:N} \mid \hat{f}_{1:N}, N, \fcut) \notag  \\
    &\propto \pi(\theta) \, P(N \mid \theta, \fcut) 
    \, \prod_{i=1}^{N} P(f_i, \hat{f}_i \mid \theta, \fcut, S) \notag \\
    &\propto \pi(\theta)\frac{\Nbar^{N}}{N!}\, \mathrm{e}^{-\Nbar} N! \prod_{i=1}^{N} \frac{\psi(f_i, \theta)\, P(\hat{f}_i \mid f_i)\, \Theta(\hat{f}_i - \fcut)}{\Nbar} \notag \\
    &\propto \pi(\theta) \, \mathrm{e}^{-\Nbar} \,\prod_{i=1}^{N} \psi(f_i, \theta)\, P(\hat{f}_i \mid f_i),
\label{eqn:full posterior}
\end{align}
where we have used Equations~\eqref{eqn:N draw} and \eqref{eqn:joint_draw}, \(\pi(\theta)\) is the prior on the population-level parameters, and \(\Theta(\hat{f}_i - \fcut) = 1\) for all \(N\) selected stars. The product \(\psi(f_i, \theta)\, P(\hat{f}_i \mid f_i)\) acts as a noise-convolved likelihood, similar to the photometric error smoothing in \cite{makarov_tip_2006}. The hierarchical Bayesian formalism used here generalises that approach: per-star uncertainties can be incorporated naturally, and stars with higher noise are downweighted in the inference. Equation~\eqref{eqn:full posterior} underpins the remainder of the paper: it is used as the basis for uncertainty forecasting in Section~\ref{sec:error forecast} and is the target for posterior sampling in Section~\ref{sec:sampling}.

\section{Uncertainty Forecast}
\label{sec:error forecast}

We expect the precision of a TRGB measurement to be governed by the photometric noise, the stellar density, and the level of AGB contamination in the field. Sampling the posterior for a single realisation as described in Section~\ref{sec:sampling} gives only one estimate of the uncertainty on \(\fTRGB\), \(\sigmaftrgb\), which is often dominated by Poisson noise. A closed-form forecast for \(\sigmaftrgb\) reveals distinct regimes that are independent of realisation. This informs the survey-design choices of Section~\ref{sec:Survey}.

For uncertainty forecasts, we exploit the fact that it is only the form of the counts close to the tip flux that is important, allowing us to adopt a simple LF and noise model. Taking \(a = b = 0\) in Equation~\eqref{eqn:full LF} yields the piecewise-constant LF
\begin{equation}  \psi(f,\theta) =
\begin{cases}
    \rhobelow & 0 < f \leq \fTRGB\\
    \rhoabove & \fTRGB < f \leq \fmax,
\end{cases}
\label{eqn: step LF}
\end{equation}
where \(\rhobelow > \rhoabove\) and the parameter list is simplified to \(\theta=(\fTRGB, \rhobelow,\rhoabove)\). The strength of the break is again usefully characterised by \(r = \rhoabove/\rhobelow\) as in Equation~\eqref{eqn:r definition}.

Similarly, we take the photometric noise, \(\sigma\), to be constant across the tip region, since most of the information about the tip is provided by stars within \(\sim\!\sigma\) of \(\fTRGB\), and even for the real data (Section~\ref{sec:Real data}) the photometric uncertainties are the same to within a few per cent. We assume the measurement noise is normally distributed so that \(P(\hat{f} \mid f) = \mathcal{N}(\hat{f}; f, \sigma^2)\).

With uniform improper priors on the three parameters in this model, defined by the density
\begin{equation}
\label{eqn:flat prior}
\pi(\theta) 
= \pi(\fTRGB, \rhobelow, \rhoabove)
\propto 
\Theta(\fTRGB) \, 
\Theta(\rhobelow) \, 
\Theta(\rhoabove),
\end{equation}
it is then possible to obtain relatively simple expressions for the uncertainty on \(\fTRGB\) for both the idealised case of no AGB (Section~\ref{sec: No AGB}) and then the more realistic case where AGB contamination is a significant effect (Section~\ref{sec:TRGB and AGB}).

\subsection{No AGB}
\label{sec: No AGB}

The simplest model for examining the uncertainty of the tip measurement assumes no AGB stars are present, with the RGB LF constant in flux. This idealisation is likely to yield the best possible measurement of \(\fTRGB\). Therefore, we begin with this model to provide a benchmark against which the more realistic cases that follow can be compared. There are two distinct sources of uncertainty: finite-number effects and photometric noise. The former dominate when the photometric noise is small compared with the mean spacing between stars at the tip; otherwise, the noise dominates, with a transition between the two regimes. We deal with these two scenarios in turn.

\subsubsection{Zero-noise limit}
\label{sec:no noise no agb}
We analyse \(\sigmaftrgb\) in the no-AGB case, where \(\rhoabove=0\). We denote the maximum measured flux value as \(\max(\hat{f}_{1:N})\equiv \fThat\) because it is the natural estimator of \(\fTRGB\). It is also a naive one, which will fail for sufficiently large noise and density. In the zero-noise limit, however, \(\fThat\) will be an increasingly good estimator of \(\fTRGB\) with increasing \(\rhobelow\).

In this regime, the posterior density from Equation~\eqref{eqn:full posterior} simplifies to 
\begin{equation}
    P(\theta \mid \hat{f}_{1:N}, N)
    \propto \pi(\theta) \, \mathrm{e}^{-\Nbar} \,\prod_{i=1}^{N} \psi(\hat{f}_i, \theta),
    \label{eqn:no noise general posterior}
\end{equation}
where 
\begin{equation}
    \Nbar = \int_{0}^{\infty} \mathrm{d} \hat{f} \, \psi(\hat{f},\theta) = \rhobelow \,\fTRGB.
\end{equation}
Using the uniform prior from Equation~\eqref{eqn:flat prior}, this becomes 
\begin{equation}
P(\theta \mid \hat{f}_{1:N}, N) \propto
    \Theta(\fTRGB-\fThat) \,\Theta(\rhobelow) \, \exp(-\rhobelow\fTRGB)\,\rhobelow^N,
\end{equation}
which is proper provided \(N \geq 1\).

Integrating over \(\rhobelow\) and normalising, the marginal posterior of \(\fTRGB\) is a Pareto distribution with shape parameter \(N\) and scale parameter \(\fThat\), with density
\begin{equation}
P(\fTRGB \mid \hat{f}_{1:N}, N) =
    \Theta(\fTRGB-\fThat)\,\frac{N}{\fThat}\left(\frac{\fTRGB}{\fThat}\right)^{-(N+1)},
    \label{eqn: no noise no agb pareto posterior}
\end{equation}
where the equality holds for \(\fmax\gg\fThat\). For large \(N\), \(N/\fThat\) becomes a good estimator for \(\rhobelow\), leaving
\begin{equation}
    \frac{\fThat}{N} \simeq \frac{\fTtrue}{N} \simeq \frac{1}{\rhobelow},
\label{eqn:fhat substitution}
\end{equation}
where \(\fTtrue\) is the true tip flux. For sufficiently large \(N\), as outlined in Appendix~\ref{app:pareto distribution}, Equation~\eqref{eqn: no noise no agb pareto posterior} approaches the distribution
\begin{equation}
P(\fTRGB \mid \hat{f}_{1:N}, N) \propto \Theta(\fTRGB-\fThat)\,\exp(-\rhobelow\fTRGB),
\label{eqn:no noise no AGB exponential}
\end{equation}
which is a shifted exponential distribution with parameter \(\rhobelow\). Substituting \(\rhobelow\) for \(N/\fThat\) in this way gives the posterior width as a function of the stellar density immediately below the tip rather than the number of stars in the catalogue, a form that is more useful when forecasting uncertainties for more realistic models. This exponential distribution has a variance of \(1/\rhobelow^2\), and so the noise-free no-AGB case has the tip uncertainty
\begin{equation}
\sigmaftrgb = \frac{1}{\rhobelow}.
\label{eqn:no-noise-no-AGB-sigma}
\end{equation}
For \(\sigma = 0\), the likelihood support depends on \(\fTRGB\), violating the regularity conditions required for asymptotic normality, leaving the posterior as asymptotically exponential rather than normal. 

\subsubsection{Noise-dominated limit}

For \(\sigma > 0\), we enter a regime where the likelihood is smoothed by its convolution with the noise kernel. We impose a low-flux cut at \(\rhostar\sigma\), retaining only stars with \(\mathrm{SNR} > \rhostar\) in the catalogue.

For sufficiently large \(N\), under regularity conditions that the noise-smoothed likelihood now satisfies, the Bernstein--von Mises theorem \citep{van_der_vaart_asymptotic_2000} implies that the marginal posterior of \(\fTRGB\) is asymptotically normal with variance \(\mathcal{I}^{-1}(\fTRGB)\), the diagonal element of the inverse Fisher information matrix corresponding to \(\fTRGB\), a result discussed in more detail in Appendix~\ref{app:Fisher Information}.

Under the assumption that \(\fTRGB \gg \rhostar\sigma\), the relevant entry of the Fisher information matrix is
\begin{equation}
    \mathcal{I}(\fTRGB) = \frac{\rhobelow}{\sigma}\int_{-\infty}^{\infty} \mathrm{d} x\, \frac{\phi(x)^2}{\Phi(x)} = \frac{D\,\rhobelow}{\sigma},
\end{equation}
where \(\phi(x)\) is the standard normal PDF, \(\Phi(x)\) is the standard normal CDF, \(x = (\fTRGB-\hat{f})/\sigma\), and \(D \simeq 0.90\) is a numerical factor.

Inverting the full Fisher information matrix, we find \(\mathcal{I}^{-1}(\fTRGB) \simeq \sigma/(D\,\rhobelow)\), implying the marginalised uncertainty on the TRGB flux is
\begin{equation}
    \sigmaftrgb \simeq \sqrt{\frac{\sigma}{D \,\rhobelow}}
    \label{eqn:noisy-no-AGB-sigma}
\end{equation}
for \(\sigma \ll \fTRGB\). The \(\sqrt{\sigma}\) dependence can be explained by recasting Equation~\eqref{eqn:noisy-no-AGB-sigma} as \(\sigmaftrgb=\sigma/\sqrt{D\,\neff}\), where \(\neff\equiv\sigma\,\rhobelow\) is the expected number of RGB stars in a flux interval of width \(\sigma\) at the tip. Whilst reducing \(\sigma\) provides a more precise tip measurement (naively giving \(\sigmaftrgb \propto \sigma\)), this also reduces the number of stars that carry information about where the tip is; these two effects partially cancel, leaving a \(\sqrt{\sigma}\) dependence. 

\begin{figure*}
    \centering
    \includegraphics[width=\textwidth]{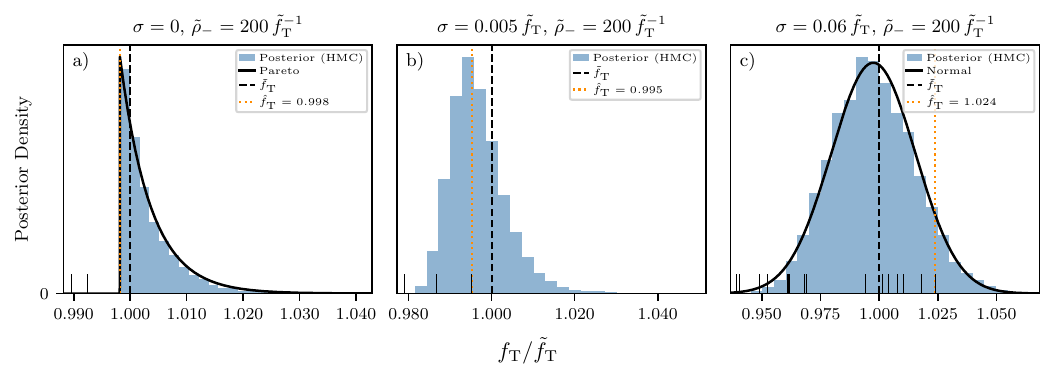}
    \caption{No-AGB (\(\rhoabove = 0\)) marginal posteriors of \(\fTRGB\) under a flat LF for three photometric uncertainties \(\sigma = 0, 0.005, 0.06 \,\fTtrue\). The density is \(\rhobelowtrue = 200\,\fTtrue^{-1}\) for all realisations, with a faint-end flux cut at \(\fcut = 5\sigma\) (\(\mathrm{SNR} = 5\)). In a), the closed-form Pareto posterior is overplotted; b) illustrates the transitional regime between the number-dominated and noise-dominated regimes; and in c), the normal distribution prediction is overplotted, centred at the posterior mean. The vertical marks at the base of each panel show the individual measured fluxes.}
    \label{fig:no-AGB-triptych-sigma}
\end{figure*}

\begin{figure*}
    \centering
    \includegraphics[width=\textwidth]{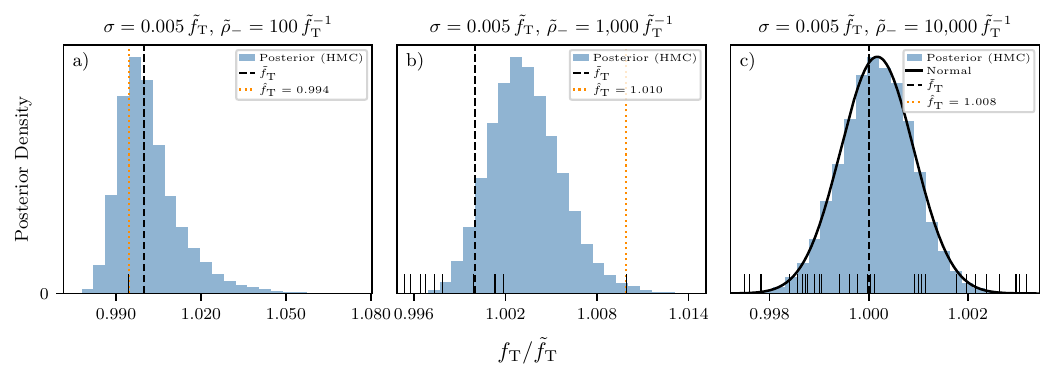}
    \caption{No-AGB (\(\rhoabove = 0\)) marginal posteriors of \(\fTRGB\) under a flat LF for three densities \(\rhobelowtrue = 100, 1000, 10{,}000\,\fTtrue^{-1}\), with \(\sigma = 0.005 \,\fTtrue\) for all realisations, and a faint-end flux cut at \(\fcut = 5\sigma\) (\(\mathrm{SNR} = 5\)). The vertical marks at the base of each panel show the individual measured fluxes. In c), the normal distribution prediction is overplotted, centred at the posterior mean, and \(\fThat\) lies beyond the plotted range.}
    \label{fig:no-AGB-triptych-density}
\end{figure*}

The marginal posterior of \(\fTRGB\) transitions between two regimes depending on the relative magnitudes of photometric noise \(\sigma\) and stellar density \(\rhobelow\). Equating the two predictions for \(\sigmaftrgb\), Equations~\eqref{eqn:no-noise-no-AGB-sigma} and \eqref{eqn:noisy-no-AGB-sigma}, implies that the transition between these two regimes occurs at \(\neff =D\simeq0.9\), \textit{i.e.}\@ when there is approximately one star within \(\sigma\) of the tip. 

\begin{figure*}
    \centering
    \includegraphics[width=\textwidth]{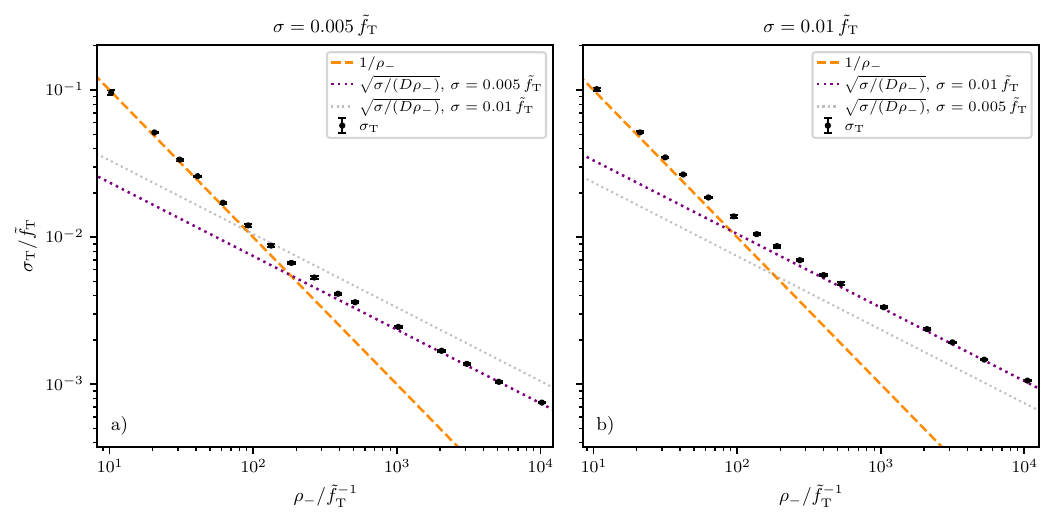}
    \caption{Posterior width \(\sigmaftrgb\), averaged over 10 realisations, plotted against \(\rhobelow\) in the no-AGB (\(\rhoabove = 0\)) case. The number- (\(1/\rhobelow\)) and noise-dominated (\(\sqrt{\sigma/[D\,\rhobelow]}\)) asymptotes are plotted as dashed orange and dotted purple lines, respectively. The transition between regimes occurs at \(\neff=D \simeq 0.9\): approximately \(180\,\fTtrue^{-1}\) for \(\sigma = 0.005\,\fTtrue\) in a) and \(90\,\fTtrue^{-1}\) for \(\sigma = 0.01\,\fTtrue\) in b). For comparison, the grey dotted line shows where the other panel's noise-dominated asymptote lies.}
    \label{fig:flat_noAGB_regimes_diptych}
\end{figure*}

Figure~\ref{fig:no-AGB-triptych-sigma} illustrates this transition with increasing \(\sigma\) at fixed \(\rhobelowtrue = 200\,\fTtrue^{-1}\): the noise-free case follows a closed-form Pareto distribution, whilst at larger \(\sigma\) the posterior converges to the Bernstein--von Mises normal prediction. With \(\rhobelowtrue = 200\,\fTtrue^{-1}\), \(\neff \simeq0.9\) implies that the transition should occur at \(\sigma\simeq 0.005\,\fTtrue\). For larger \(\sigma\), \(\fThat\) contributes less information about \(\fTRGB\). For \(\sigma = 0\), it acts as a hard lower bound on the posterior of \(\fTRGB\), whereas for increasing \(\sigma\), \(\fThat\) moves to higher flux, through the posterior mean, and eventually lies well above the mean for large \(\neff\). 

Figure~\ref{fig:no-AGB-triptych-density} shows that the same convergence occurs with increasing \(\rhobelowtrue\) at fixed \(\sigma = 0.005\,\fTtrue\). Figure~\ref{fig:flat_noAGB_regimes_diptych} shows the standard deviation on the marginal posterior of \(\fTRGB\) for increasing \(\rhobelow\) at fixed \(\sigma\) and demonstrates that a smooth shift between the two regimes occurs at the predicted values of \(\rhobelow\). As \(\sigma\) grows, the transition occurs at smaller densities. 

\subsection{TRGB and AGB}
\label{sec:TRGB and AGB}

Having established the no-AGB baseline, we now add the AGB contaminant population to the LF through \(\rhoabove>0\), so that the density ratio \(r\) enters the forecast. As with the no-AGB case, we start by examining \(\sigmaftrgb\) in the noise-free limit, before adding photometric noise.

\subsubsection{Zero-noise limit}

The \(\sigma = 0\) posterior from Equation~\eqref{eqn:no noise general posterior} can be rewritten for a given \(\fTRGB\) in terms of the \(\Nbelow\) stars with \(f\leq\fTRGB\), and \(\Nabove\) stars with \(f>\fTRGB\) (with \(N = \Nbelow + \Nabove\)). Adopting the uniform prior from Equation~\eqref{eqn:flat prior}, the posterior simplifies to 
\begin{equation}
\label{eqn:no noise agb rgb posterior}
    P(\theta \mid \hat{f}_{1:N}, N) \propto \pi(\theta) \, \mathrm{e}^{-\Nbar} \,\rhobelow^{\Nbelow}\,\rhoabove^{\Nabove},
\end{equation}
where 
\begin{equation}
    \Nbar = \rhobelow\,\fTRGB + \rhoabove(\fmax-\fTRGB).
\end{equation}
From this, we can see that for a fixed trial \(\theta=(\fTRGB,\,\rhobelow,\,\rhoabove)\), the sufficient statistics are \((\Nbelow,\Nabove)\), and the individual flux values themselves do not inform the likelihood beyond which side of the tip they fall on.

Integrating Equation~\eqref{eqn:no noise agb rgb posterior} over \(\rhobelow\) and \(\rhoabove\), we obtain the marginal posterior of \(\fTRGB\) as
\begin{equation}
    P(\fTRGB \mid \hat{f}_{1:N}, N) \propto 
    \frac{\Nbelow!\,\Nabove!}
    {\fTRGB^{\Nbelow+1}\,
    (\fmax - \fTRGB)^{\Nabove+1}}.
\label{eqn:TRGB and AGB no noise}
\end{equation}
There is no closed-form expression for \(\sigmaftrgb\) in this \(\sigma=0\) case. Equation~\eqref{eqn:TRGB and AGB no noise} depends on \(\fTRGB\) both explicitly and through \(\Nbelow\) and \(\Nabove\), which are step functions of \(\fTRGB\) with unit jumps at each of the \(N\) observed fluxes, so the marginal posterior is discontinuous there. These discontinuities are present regardless of the underlying densities \(\rhobelow\) and \(\rhoabove\). 

\begin{figure*}
    \centering
    \includegraphics[width=\textwidth]{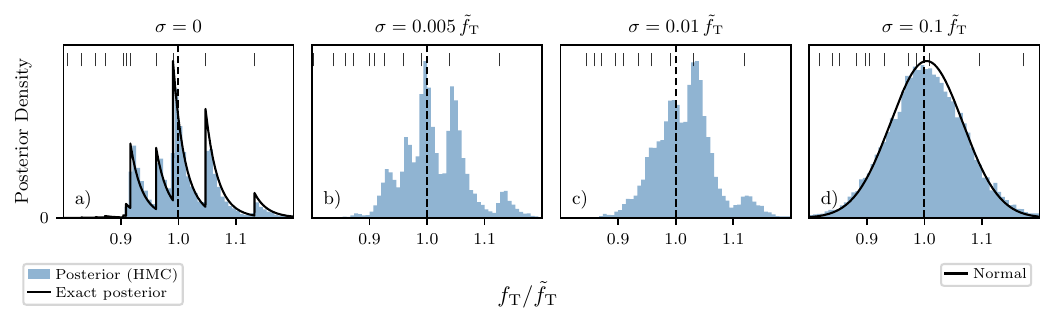}
    \caption{Marginal posterior of \(\fTRGB\) in the \(\rhoabove>0\) case, for the same simulated stellar catalogue observed at four different photometric noise levels, with \(\rhobelowtrue=50\,\fTtrue^{-1}\) and \(\rhoabovetrue=5\,\fTtrue^{-1}\). The four panels share the same true fluxes; the measured fluxes are shown as vertical marks. The exact posterior of the noise-free (\(\sigma=0\)) case (Equation~\ref{eqn:TRGB and AGB no noise}) is overplotted in a), and the normal prediction of Equation~\eqref{eqn:bvm_step} is overplotted in d), centred at the posterior mean.}
    \label{fig:exact_step_quad}
\end{figure*}

\subsubsection{Noise-dominated limit}

The smoothing from measurement noise removes the above discontinuity; for sufficiently large \(\sigma\), we enter a regime where the Bernstein--von Mises theorem again applies, and so the marginal posterior of \(\fTRGB\) is asymptotically normal. Inverting the full Fisher information matrix calculated using Equation~\eqref{eqn:fisher-information-PPP} with both \(\rhobelow\) and \(\rhoabove >0\), we find that the entry of the inverse matrix corresponding to \(\fTRGB\) is
\begin{equation}
    \mathcal{I}^{-1}(\fTRGB) \simeq \frac{\sigma}{D(r)\,\rhobelow\,(1-r)^2},
\end{equation}
where
\begin{equation}
    D(r) = \int_{-\infty}^{\infty} \mathrm{d} x\, \frac{\phi(x)^2}{r+(1-r)\,\Phi(x)},
\end{equation}
yielding
\begin{equation}
    \sigmaftrgb \simeq \sqrt{\frac{\sigma}{D(r)\,\rhobelow\,(1-r)^2}}
\label{eqn:bvm_step}
\end{equation}
for \(\sigma \ll \fTRGB\). The same explanation for the \(\sqrt{\sigma}\) dependence from Equation~\eqref{eqn:noisy-no-AGB-sigma} holds here too: reducing the photometric error sharpens the edge but also reduces the number of stars carrying information about it. 

Figure~\ref{fig:exact_step_quad} shows the transition from the \(\sigma=0\) case to the regime where the Bernstein--von Mises normal approximation holds. Each panel of Figure~\ref{fig:exact_step_quad} uses the same set of true fluxes and the same noise vector, scaled to each uncertainty level. For larger \(\sigma\), the peaks in the discontinuous posterior distribution become smoothed out, and the resulting posterior approaches the predicted normal. Equation~\eqref{eqn:bvm_step} gives only the width of that normal, so the overplotted curve is centred at the posterior mean. 

\subsection{Survey design}
\label{sec:Survey}

One way in which the above uncertainty forecasts can be used is to assess observational strategies. In different ways, a survey can control \(\rhobelow\), \(\sigma\), and \(r\). The most intuitive is \(\rhobelow\), which (at constant \(r\)) is a function of the total number of stars in a single galaxy that a survey images. A survey that opts to sample twice the number of fields (with all else being equal) will reduce \(\sigmaftrgb\) by a factor of \(\sqrt{2}\). The same result also follows from surveying a denser galaxy or increasing the angular size of each field. 

A weaker lever is the photometric uncertainty, \(\sigma\), which falls with exposure time as \(\sigma \propto T^{-1/2}\) for shot-noise-limited photometry, giving \(\sigmaftrgb\propto T^{-1/4}\). For \(J\) fields each of exposure time \(T\), \(\rhobelow \propto J\) and the total exposure time is \(t = JT\), giving \(\sigmaftrgb \propto J^{-1/2}\, T^{-1/4} = J^{-1/4}\,t^{-1/4}\): at fixed total exposure time, a wide, shallow survey is
preferable to a narrow, deep one. This only holds for \(\sigma\ll\fTRGB\), since the flux at a fixed SNR cut rises and eventually reaches the tip for high \(\sigma\).

\begin{figure}
    \centering
    \includegraphics[width=\columnwidth]{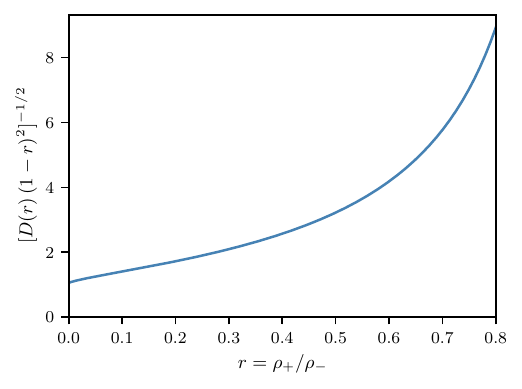}
    \caption{Inflation of the forecast tip uncertainty with AGB contamination, writing Equation~\eqref{eqn:bvm_step} as \(\sigmaftrgb = [D(r)(1-r)^2]^{-1/2}\sqrt{\sigma/\rhobelow}\). We plot the function for \(r\leq0.8\) because it diverges as \(r\rightarrow1\), and the asymptotic normality assumed in deriving Equation~\eqref{eqn:bvm_step} degrades well before that, as the tip becomes poorly identified (Section~\ref{sec:Sensitivity to AGB Contamination}).}
    \label{fig:r_inflation}
\end{figure}

The steepest dependence is on \(r\), shown in Figure~\ref{fig:r_inflation}. Although \(r\) is a property of the stellar population, a survey can be designed to minimise it. The ratio is set by the colour-band selection and by field placement, with outer-disc and halo fields carrying lower AGB fractions \citep{freedman_status_2025}. One caveat with using fields further from the centre of the galaxy is that these regions are also less dense, causing \(\rhobelow\) to fall, so imaging them is only beneficial if the reduction in \(r\) outweighs the loss in \(\rhobelow\).

\section{Sampling the Full Posterior Distribution}
\label{sec:sampling}

The joint posterior of Equation~\eqref{eqn:full posterior}, on the population-level parameters, \(\theta = (\fTRGB, a, b, \rhobelow, \rhoabove)\), and the \(N\) latent true fluxes, \(f_{1:N}\), has no closed form, so we must draw samples from it instead. Any marginal distribution of interest (most obviously \(\fTRGB\)) is then obtained simply by discarding the samples of the remaining parameters. The sampling task is numerically challenging, and we found it was necessary to both reparametrise the problem (Section~\ref{sec:reparam}) and make careful prior choices (Section~\ref{sec:priors}) to enable efficient sampling (Section~\ref{sec:hmc settings}).

\subsection{Reparametrisation}
\label{sec:reparam}
The parametrisation of the LF in Equation~\eqref{eqn:full LF} exhibits a degeneracy between the scaling \(\rhobelow\) and \(\fTRGB\). The reparametrisation
\begin{equation}
    \psi(f,\theta) =
\begin{cases}
    \alpha f^{-a} & \fmin \leq f\leq\fTRGB\\
    \beta f^{-b} & f>\fTRGB,
\end{cases}
\label{eqn:reparametrised LF}
\end{equation}
with \(\alpha = \rhobelow\, \fTRGB^{a}\) and \(\beta = \rhoabove\, \fTRGB^{b}\), removes \(\fTRGB\) from the amplitude, constraining it only through the location of the discontinuity, reducing its correlation with the amplitude, and improving sampler stability. Priors are placed on the sampled parameters \(\log\alpha,\,\log\beta,\,\log \fTRGB,\, a,\) and \(b\) (Section~\ref{sec:priors}), together with the \(N\) latent true fluxes.

\subsection{Priors}
\label{sec:priors}
We adopt scale-invariant (log-uniform) priors on the amplitudes and on \(\fTRGB\), and uniform priors on the slopes:
\begin{equation}
    \pi(\log\alpha,\,\log\beta,\,\log \fTRGB,\,a,\,b) \propto 1,
    \label{eqn:priors}
\end{equation}
within the bounds \(\log\alpha,\,\log\beta \in [\log 10^{-2},\,\log 10^{5}]\), \(\log \fTRGB \in [\log (\rhostar\sigma),\,\log \fmax]\), \(a \in [0.01,\, 5]\), and \(b \in [1.01,\, 5]\), and zero otherwise. Here, \(\fmax\) is the maximum flux value that the AGB is simulated to, well above \(\fTRGB\). A uniform prior on \(\log \fTRGB\) is equivalent to a uniform prior on the magnitude of the TRGB, \(\mTRGB\). The bounds on \(a\) and \(b\) are chosen to exclude unphysical slopes.

\subsection{HMC sampler settings}
\label{sec:hmc settings}

We sample from the \((N+5)\)-dimensional joint posterior using Hamiltonian Monte Carlo (HMC; \citealt{neal_mcmc_2011}) as implemented in \texttt{Stan} (Version~2.36; \citealt{stan_development_team_stan_2026}), using the \texttt{Python} interface \texttt{CmdStanPy} (Version~1.2.5; \citealt{stan_development_team_cmdstanpy_2025}).

To ensure that the LF is differentiable for HMC, we smooth the discontinuity at \(\fTRGB\) with a logistic transition in \(\log{f}\) of width \(s_{\mathrm{LF}}=0.001\). This smoothing is an order of magnitude smaller than the photometric noise (see Section~\ref{sec:coverage} for the simulation parameters) and is added purely for sampler stability. 

The integral for the expected number of stars in the catalogue, Equation~\eqref{eqn:N_bar}, does not have a closed-form solution. For the TRGB, we evaluate the integral using Simpson's rule on a uniform grid in \(\log f\) with 200 subintervals over a range \([\log (\rhostar\sigma - 5\sigma), \log \fTRGB]\). The lower cut is made for computational efficiency: below \(\rhostar\sigma - 5\sigma\), \(P(S \mid f, \fcut)<3\times10^{-7}\), and the fractional error in \(\Nbar\) from this truncation is of order \(10^{-7}\), five orders of magnitude below the Poisson noise \(1/\sqrt{N}\) for a typical RGB LF.

For the AGB, we assume that the cut is far enough from the tip that \(P(S\mid f, \fcut) = 1\) for \(f>\fTRGB\), leaving the AGB integral with a closed form.

Each realisation's \((N+5)\)-dimensional joint posterior is sampled using four chains, 2000 warm-up iterations, and 4000 sampling iterations, with default Stan sampler settings. We initialise the chains near the true values with a small jitter for synthetic runs. For real-data runs, the chains are initialised at estimates of the parameter values.

This yields an effective sample size of \(\sim\!5000\) (from 16{,}000 samples) in \(\sim\!10\) minutes on an Apple M3 Pro (11-core CPU, 36 GB RAM) for \(N\sim\!3000\) stars. We verify convergence using the \cite{gelman_inference_1992} statistic, requiring \(\hat{R} < 1.01\) for all parameters, with no divergences.

\section{Analysis of Simulated Data}
\label{sec:coverage}

We perform fixed-value calibration tests to verify the unbiased recovery of fixed true values for different realisations of the population. We generate 300 datasets using the generation pipeline described in Section~\ref{sec:Generative Model}, choosing true values, \(\tilde\theta\), for the population-level parameters that align with NGC~4258 fields observed in TRGB survey catalogues (\textit{e.g.}\@ Extragalactic Distance Database; \citealt{tully_extragalactic_2009, anand_extragalactic_2021}). The fixed values chosen for the population-level parameters are \(\tilde a = 2.8\), \(\tilde b = 3.5\), \(\rhobelowtrue = 1400\,\fTtrue^{-1}\), \(\rhoabovetrue = 600\,\fTtrue^{-1}\), with fluxes in units of the true tip flux, \(\fTtrue\). The photometric uncertainty is taken to be constant in flux, \(\sigma = 0.024\,\fTtrue\). We impose a per-star SNR cut, \(\hat{f}/\sigma>\rhostar=15\). The true flux distribution is bounded by \(\fmin=0.04\,\fTtrue\) and \(\fmax=4\times10^5\,\fTtrue\). With these choices, \(\Nbar_{\mathrm{LF}}=2.5\times10^5\), and (after applying the photometric noise and SNR cut) \(\sim\!4400\) stars enter the catalogue.

\begin{figure}
    \centering
    \includegraphics[width=\columnwidth]{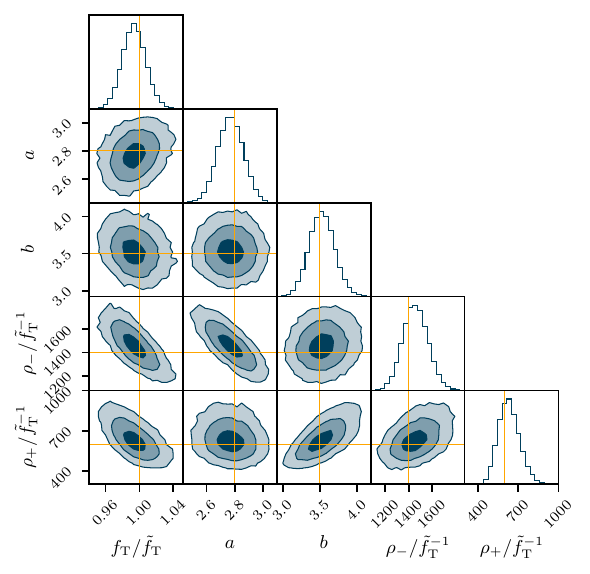}
    \caption{Corner plot showing the marginal posteriors of the five population-level parameters for a single representative synthetic realisation (the true values, marked with orange lines and their intersections, are listed in Section~\ref{sec:coverage}). Filled contours enclose 39.3\%, 86.5\%, and 98.9\% of the posterior probability in each panel.}
    \label{fig:simulated corner plot}
\end{figure}

\begin{figure*}
    \centering
    \includegraphics[width=\textwidth]{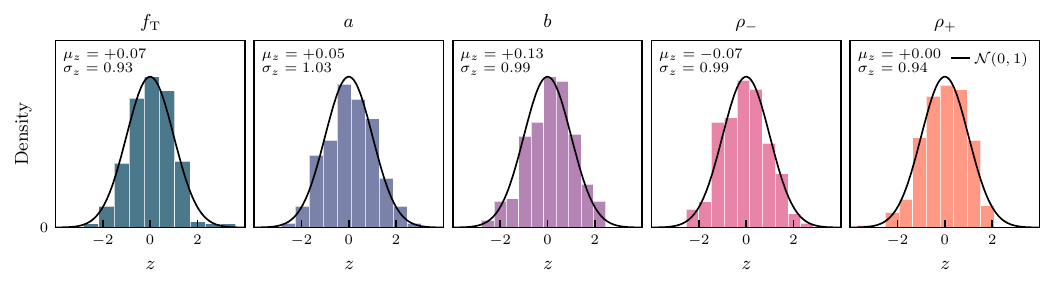}
    \caption{Distribution of posterior \(z\)-scores of population-level parameters for 300 simulations with fixed true values listed in Section~\ref{sec:coverage}. The solid black curve is \(\mathcal{N}(0,1)\), and the mean, \(\mu_z\), and standard deviation, \(\sigma_z\), are annotated in each panel.}
    \label{fig:simulated z score plot}
\end{figure*}

An example of the resultant posterior distributions is shown in Figure~\ref{fig:simulated corner plot}. The correlation between the shape and density parameter pairs \((a, \rhobelow)\) and \((b, \rhoabove)\) is expected: simultaneously increasing (decreasing) the slopes and decreasing (increasing) the amplitudes leaves the likelihood largely unchanged. In this realisation, \(\fTRGB = 0.992 \pm 0.012 \,\fTtrue\).

For each realisation we measure the mean, \(\bar{\theta}\), and standard deviation, \({\sigma}_{\theta}\), of the marginal posterior of each parameter and calculate the \(z\)-score,
\(z= (\bar{\theta} - \tilde{\theta}) / \sigma_\theta\),
which measures, in units of the posterior width, the distance between the posterior mean and the truth. If the posterior is correctly calibrated at the chosen true values, \(z\) is distributed as \(\mathcal{N}(0,1)\). A non-zero mean indicates bias, whilst a width greater (less) than one indicates overconfidence (underconfidence). 

As shown in Figure~\ref{fig:simulated z score plot}, the \(z\)-scores of all five parameters are consistent with \(\mathcal{N}(0,1)\): their means lie within 0.13 of zero and their standard deviations within 0.07 of one, against Monte Carlo uncertainties of 0.06 on the means and 0.04 on the standard deviations for 300 realisations. The largest mean offset is on \(b\), which is the least-constrained parameter. The inference is unbiased and well calibrated at the chosen truth. We use fixed-value calibration rather than full simulation-based calibration (SBC; \citealt{talts_validating_2020}), as the uniformity of the SBC rank statistics holds only if the entire prior is sampled, so the test cannot be restricted to a computationally feasible subset. The priors we use (Section~\ref{sec:priors}) are deliberately wide, chosen to let the data constrain the parameters rather than to describe a real field. Sampling from the priors gives fields with unphysical forms, where \(\rhoabove>\rhobelow\), and catalogue sizes that range from too few stars to constrain a tip to far more than any observed field, for which the \((N+5)\)-dimensional posterior would be intractable. Fixed-value calibration instead tests the inference at a single, physically motivated point, at the cost that calibration is not guaranteed elsewhere in parameter space.

\subsection{Sensitivity to AGB contamination}
\label{sec:Sensitivity to AGB Contamination}

\begin{figure}
    \centering
    \includegraphics[width=\columnwidth]{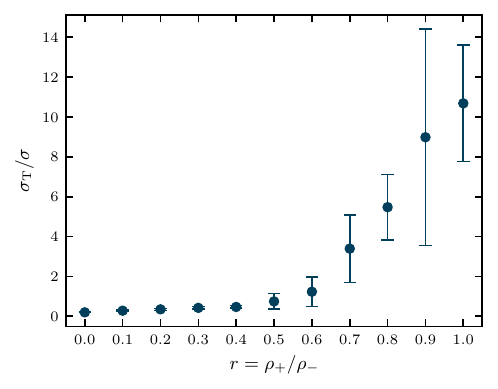}
    \caption{The mean and standard deviation of \(\sigmaftrgb\), the standard deviation of the marginal posterior of \(\fTRGB\), plotted for up to 10 realisations at each \(r=\rhoabove/\rhobelow\) value. Realisations failing the convergence criteria of Section~\ref{sec:hmc settings} are excluded, which removes one realisation at \(r=0.7\) and two at \(r=1\).}
    \label{fig:r dependence plot}
\end{figure}

We examine the ability of the model to infer \(\fTRGB\) in cases of high contamination. For \(r\) as defined in Equation~\eqref{eqn:r definition}, we test the model with \(r \in \{0, 0.1, \ldots, 1\}\). If \(r=0\), there is no AGB contamination, and if \(r=1\), the LF at the TRGB is continuous: the TRGB becomes a change-point in slope. The density \(\rhobelowtrue\) and all the other population-level parameters are maintained from Section~\ref{sec:coverage}, and \(\rhoabovetrue\) is varied to give the chosen density ratios. For reference, using the values \(\rhobelowtrue = 1400\,\fTtrue^{-1}\) and \(\rhoabovetrue = 600\,\fTtrue^{-1}\) in Section~\ref{sec:coverage} gives \(r=0.43\). Figure~\ref{fig:r dependence plot} shows how \(\sigmaftrgb\) varies across the
realisations at each \(r\).

We see two regimes: for \(r\lesssim0.5\), \(\sigmaftrgb\) increases from \(0.2\sigma\) to \(0.75\sigma\) and has little scatter between realisations until \(r=0.5\), confirming the simpler calculation in Section~\ref{sec:TRGB and AGB}. Beyond this, the density jump at the tip is too shallow to dominate: random fluctuations in the observed fluxes create local maxima in the likelihood, leading to multimodal posteriors. In turn, \(\sigmaftrgb\) becomes larger and more variable between realisations, until the change in the LF slope is the only signal at \(r=1\). The multimodality and poor identifiability are finite-sample effects: for increasing \(N\), we expect the \(r\) value that defines the change in regimes to increase, as the discontinuity becomes better defined, and for large enough \(N\), the slope change alone should identify \(\fTRGB\) reliably, even at \(r=1\). However, for the catalogue sizes considered here, fields with \(r\gtrsim0.5\) should not be used for a tip measurement. This is a stronger reason to prefer fields with low AGB contamination (\textit{i.e.}\@ in the outskirts of galaxies) than the inflation of \(\sigmaftrgb\) forecast in Section~\ref{sec:Survey}: at high \(r\), the tip may not be identified at all, rather than identified imprecisely. 

\section{Heteroscedastic Photometric Uncertainty}
\label{sec:Heteroscedastic Photometric Uncertainty}
Real photometric catalogues exhibit heteroscedastic\footnote{A noise model is heteroscedastic if its variance differs between observations, as opposed to homoscedastic noise, whose variance is the same for all observations. Here, the variance of each star's flux measurement depends on that star's flux, and there is also the possibility of data with different background levels.} and flux-dependent photometric uncertainties. We extend our photometric noise model to account for this (Section~\ref{sec:Photometric noise model}) and use it to show how not accounting for these effects can bias the TRGB inference (Section~\ref{sec:noise sim}). We then provide a practical approach for applying this to real catalogue-level data (Section~\ref{sec:Application to catalogue-level data}).

\subsection{Photometric noise model}
\label{sec:Photometric noise model}

The photometric uncertainty in flux can be modelled as the combination of a constant read/background noise term and Poissonian photon shot noise. The shot noise is set by the zero-point flux of the survey instrumentation, \(f_\mathrm{z}\) --- the flux of a source that produces, on average, one count in time \(T_\mathrm{z} = 1\)~s on the detector --- together with the exposure time, \(T\), of each observation, giving 
\begin{equation}
    \sigma^2(f) = \sigma_0^2 + f\,f_{\mathrm{z}}\,\frac{T_{\mathrm{z}}}{T} \equiv \sigma_0^2 + Cf,
\end{equation}
where \(\sigma_0\) is the flux-independent (read/background noise) term and \(C =f_{\mathrm{z}} T_{\mathrm{z}}/T\) has units of flux.

With this definition of \(C\), we include this prescription of the uncertainty in our likelihood. The uncertainty appears in both the Poisson expected-number factor, \(\Nbar\), and the measurement model in the object-level product. 

We account for the Poisson contribution to the measurement noise by adopting \(P(\hat{f}_i | f_i) = \mathcal{N}(\hat{f}_i; f_i, \sigma_0^2 + Cf_i)\) in Equation~\eqref{eqn:full posterior}. The measurement now depends on the true flux of each star, which is concurrently inferred. 

To include the heteroscedastic noise in \(\Nbar\), we first determine the constant flux cut, \(\fcut\), which is the positive solution of the equation \(\fcut=\rhostar\sigma(\fcut)\), giving
\begin{equation}
    \fcut = \frac{C\,\rhostar^2+\sqrt{C^2\,\rhostar^4+4\,\rhostar^2\,\sigma_0^2}}{2}.
\label{eqn:fcut}
\end{equation}
In the limit \(C\to0\), we recover the constant-in-flux noise case, \(\fcut= \rhostar\sigma_0\).

Substituting \(P(\hat{f} \mid f) = \mathcal{N}(\hat{f}; f, \sigma_0^2 + Cf)\) into Equation~\eqref{eqn:P(S|f)},
\begin{equation}
    P(S \mid f, \fcut) = 1 - \Phi\left(\frac{\fcut - f}{\sqrt{\sigma_0^2 + Cf}}\right),
\end{equation}
where \(\Phi(x)\) is the standard normal CDF, and the expected catalogue size is
\begin{equation}
    \Nbar(\theta, \fcut) = \int_0^\infty \mathrm{d}f\, \psi(f, \theta)\, P(S \mid f, \fcut).
\end{equation}

\subsection{Application to simulated data}
\label{sec:noise sim}

Following the generative model in Section~\ref{sec:Generative Model}, with \(\sigma^2(f) = \sigma_0^2 + Cf\), we generate data with heteroscedastic noise (\(\tilde\sigma_0 = 0.024\,\fTtrue\), \(\tilde C = 6.4\times10^{-4}\,\fTtrue\)) and test the effect of three different prescriptions of \(P(S \mid f, \fcut)\), summarised in Table~\ref{tab:nbar_comparison}. For brevity, we define \(S(f, \sigma)\equiv 1 - \Phi[(\fcut - f)/\sigma]\), which gives the probability that a star with true flux \(f\) and photometric uncertainty \(\sigma\) is selected. Model~1 uses the noise-free hard-cut approximation to \(\Nbar\). Models~2 and 3 use smooth selection functions with different noise prescriptions. Model~2 uses \(\sigma(\fcut)\) to see the effect of using a constant flux uncertainty in the selection function. Model~3, which uses the correct form of \(\sigma(f)\), matches the data-generation process and is taken as the truth.

\begin{table}[ht]
\renewcommand{\arraystretch}{1.9}
\centering
\caption{\(P(S \mid f, \fcut)\) models}
\label{tab:nbar_comparison}

\begin{tabular}{cc}
\toprule
Model & \(P(S \mid f, \fcut)\) \\
\midrule
1 & \(\Theta(f - \fcut)\) \\
2 & \(S(f, \sigma(\fcut))\) \\
3 & \(S(f, \sigma(f))\) \\
\bottomrule
\end{tabular}
\end{table}

\begin{figure}
    \centering
    \includegraphics[width=\columnwidth]{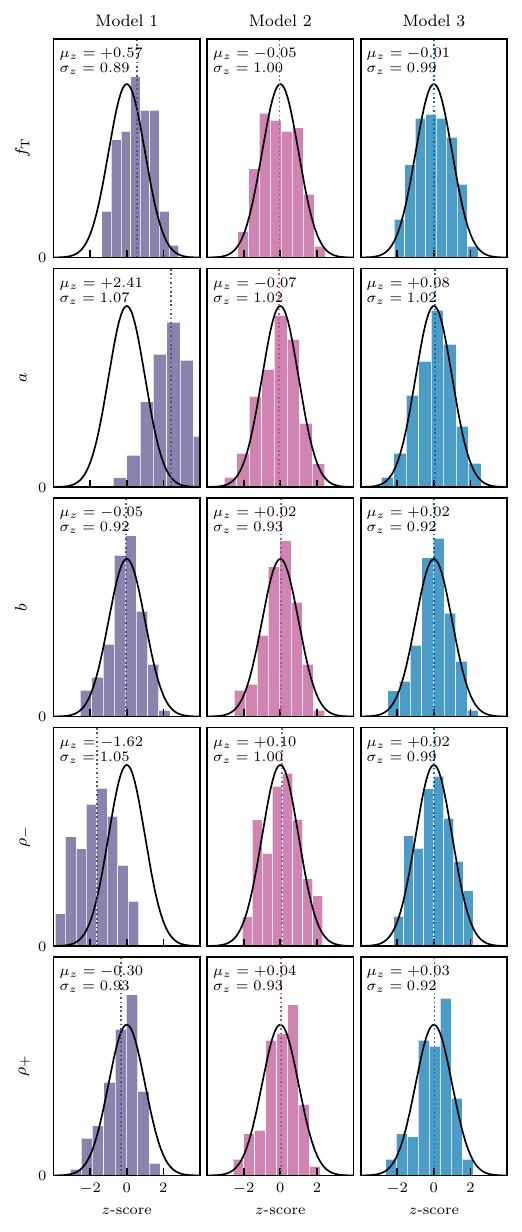}
    \caption{Distribution of posterior \(z\)-scores of population-level parameters for 200 simulations with heteroscedastic noise, at the fixed true values listed in Section~\ref{sec:coverage}, under each of the selection functions listed in Table~\ref{tab:nbar_comparison}. The solid black curve is \(\mathcal{N}(0,1)\), and the mean, \(\mu_z\), and standard deviation, \(\sigma_z\), are annotated in each panel.}
    \label{fig:nbar comparison z-scores}
\end{figure}

Histograms of the \(z\)-scores of population-level parameter posteriors from 200 realisations at the same population-level parameter truths as in Section~\ref{sec:coverage}, for each of the models, are shown in Figure~\ref{fig:nbar comparison z-scores}. Comparing the \(\fTRGB\) histograms of the three models, it is evident that correctly specifying the selection function is essential for unbiased inference of \(\fTRGB\): Model~1 shows strong bias towards higher \(\fTRGB\) values, of \(\sim\!0.5 \sigmaftrgb\). Model~2 appears well calibrated after 200 realisations. As expected, Model~3 gives well calibrated posteriors. 

The \(a\) histogram illustrates the origin of the bias in Model~1. The inference of \(a\) is strongly biased to high values (\(+1.9\sigma_a\)). This is to be expected: misspecifying the selection as a step function forces each catalogue star's inferred true flux, \(f\), to be \(>\fcut\). In the generative model, however, stars with \(f\simeq\fcut\) are scattered both in and out of the catalogue. The gradient of the RGB LF at the cut means that more stars scatter into the catalogue from below than scatter out from above (Eddington bias; \citealt{eddington_formula_1913}). The model is forced to assign these up-scattered stars to have true fluxes \(f > \fcut\), creating an artificial overdensity in the inferred true-flux distribution just above the cut. The model interprets this as a steeper RGB slope, biasing \(a\) high. We find a slight positive posterior correlation between \(a\) and \(\fTRGB\) in Figure~\ref{fig:simulated corner plot}, which carries the bias into the \(\fTRGB\) inference. As a further consequence, the steeper slope predicts more stars in the RGB, so the model biases \(\rhobelow\) low to keep \(\Nbar\) consistent with the observed \(N\) stars. Despite \(\fTRGB\) lying \(> 15\sigma(\fcut)\) above the selection threshold, the misspecified selection function biases the tip estimate by \(\sim\!0.5\sigmaftrgb\). Models~2 and 3 avoid this bias because their smooth selection functions correctly account for the scatter across the cut, thereby naturally correcting the Eddington bias. Model~2 is unbiased because the gradient of \(\sigma(f)\) is small enough at the cut: the approximation \(\sigma(f)\simeq \sigma(\fcut)\) holds well over the width of the selection window, leading to no detectable bias in any of the parameters.

\subsection{Application to real catalogue-level data}
\label{sec:Application to catalogue-level data}

With real data, the true underlying noise model is unknown: working at the catalogue level, we know only the measured fluxes, \(\hat{f}_{1:N}\), of the \(N\) stars in the catalogue and estimates of the uncertainties on each of those measured fluxes, \(\hat{\sigma}_{1:N}\). Each \(\hat{\sigma}_i\) is produced by a photometry pipeline (\textit{e.g.}\@ DAOPHOT, \citealt{stetson_daophot_1987}; DOLPHOT, \citealt{dolphin_wfpc2_2000}), which models each star as a point spread function (PSF) and fits it to the image data to extract flux and uncertainty estimates of the source. These per-star uncertainties are estimated from the observed image counts and therefore depend on the measured flux, \(\hat{f}\), rather than the unknown true flux. 

In practice, the reported \(\hat{\sigma}_{1:N}\) scatter about a smooth locus due to crowding, spatially varying sky background, PSF variability across the detector, and other field-dependent effects. We use the per-star \(\hat{\sigma}_i\) directly in the measurement model, capturing the star-to-star scatter rather than smoothing over it. This means that high-noise stars are naturally downweighted, and low-noise stars contribute more to the likelihood.

The \(\Nbar\) integral requires a parametric form of the uncertainty to be known for all true fluxes, including stars not present in the catalogue. We assume that the noise follows the same parametric form \(\sqrt{\hat{\sigma}_0^2 + \hat{C}f}\) when evaluated at true flux \(f\) as when evaluated at measured flux \(\hat{f}\). We fit the parametric form \(\sqrt{\hat{\sigma}_0^2 + \hat{C}\hat{f}}\) to the observed \((\hat{f}_i, \hat{\sigma}_i)\) distribution, giving the estimators \(\hat{\sigma}_0\) and \(\hat{C}\), from which the selection threshold \(\fcut\) and selection function \(P(S\mid f, \fcut)\) are computed.

This is incorporated into the Stan model used for the real data. In addition, we extend the reparametrisation from Equation~\eqref{eqn:reparametrised LF} to sample the density ratio across the tip, \(r\), as defined in Equation~\eqref{eqn:r definition}, so that \(\rhoabove = r\,\rhobelow\). We recover \(\beta\) using
\begin{equation}
    \beta = r\,\alpha\,\fTRGB^{b-a}. 
\end{equation}
This reparametrisation allows us to impose a downward step in stellar density in the LF at \(\fTRGB\) by bounding \(r<1\) through the choice of the prior \(\log r < 0\). We also widen the uniform priors that were used in the synthetic model of Section~\ref{sec:sampling} to \(a\in [0.01,10]\) and \(b \in [1.01, 100]\). 

We fit this model to real catalogue data in Section~\ref{sec:Real data}. 

\section{Demonstration on Real Data}
\label{sec:Real data}

\begin{table*}
\centering
\caption{Summary of NGC~4258 field TRGBs in the F814W filter. The tip magnitudes, \(\mTRGB\), and density ratios, \(r\), are posterior medians; the quoted \(\mTRGB\) uncertainties are the 16th to 84th percentiles. The observed columns are reddened; the extinction-corrected columns have Milky Way foreground and per-field internal extinction removed (see footnotes). Both ``this work'' and CATs observed tips are directly comparable, as are the two extinction-corrected tips, with the caveat that the CATs corrected tips are additionally tip--contrast standardised.}
\label{tab:ngc4258_fields}
\begin{threeparttable}
{\renewcommand{\arraystretch}{1.5}
\begin{tabular}{llrrrrrrr}
\toprule
& & & & & \multicolumn{2}{c}{Observed (reddened)} & \multicolumn{2}{c}{Extinction-corrected\tnote{f}} \\
\cmidrule(lr){6-7}\cmidrule(lr){8-9}
Field & Programme & \ch{$\rho_{*}$} & \ch{$N$} & \ch{$r$} & \ch{This work (mag)} & \ch{CATs (mag)} & \ch{This work (mag)} & \ch{CATs (mag)\tnote{g}} \\
\midrule
NGC~4258-1    & GO-9477\tnote{a}    & 24.0   & 761  & 0.472 & $25.351^{+0.024}_{-0.024}$ & $25.360 \pm 0.041$ & $25.302^{+0.024}_{-0.024}$ & $25.375 \pm 0.041$ \\
\quad(CATs second tip)  &  &  &  &  &  & $25.690 \pm 0.462$ &  & $25.619 \pm 0.462$ \\
NGC~4258-2    & GO-10399\tnote{b}   & 14.0   & 438  & 0.386 & $25.391^{+0.363}_{-0.035}$ & $25.367 \pm 0.108$ & $25.347^{+0.363}_{-0.035}$ & $25.332 \pm 0.108$ \\
NGC~4258-3    & GO-10399\tnote{b}   & 8.0    & 419  & 0.392 & $25.417^{+0.077}_{-0.047}$ & $25.334 \pm 0.130$ & $25.376^{+0.077}_{-0.047}$ & $25.301 \pm 0.130$ \\
NGC~4258-4 G1 & GO-10399\tnote{b}   & 17.0   & 692  & 0.336 & $25.312^{+0.151}_{-0.034}$ & $25.304 \pm 0.048$ & $25.269^{+0.151}_{-0.034}$ & $25.291 \pm 0.048$ \\
NGC~4258-4 G2 & GO-10399\tnote{b}   & 11.0   & 494  & 0.382 & $25.392^{+0.057}_{-0.039}$ & $25.288 \pm 0.041$ & $25.350^{+0.057}_{-0.039}$ & $25.296 \pm 0.041$ \\
NGC~4258-5    & GO-16198\tnote{c}   & 11.0   & 1381 & 0.593 & $25.501^{+0.064}_{-0.043}$ & $25.441 \pm 0.041$ & $25.455^{+0.064}_{-0.043}$ & $25.453 \pm 0.041$ \\
\quad(CATs second tip)  &  &  &  &  &  & $25.781 \pm 0.567$ &  & $25.707 \pm 0.567$ \\
NGC~4258-6    & GO-16198\tnote{c}   & 11.0   & 1238 & 0.519 & $25.489^{+0.047}_{-0.038}$ & $25.462 \pm 0.041$ & $25.445^{+0.047}_{-0.038}$ & $25.468 \pm 0.041$ \\
NGC~4258-7    & GO-16688\tnote{d}   & 27.5 & 659  & 0.303 & $25.272^{+0.018}_{-0.018}$ & $25.280 \pm 0.041$ & $25.226^{+0.018}_{-0.018}$ & $25.306 \pm 0.041$ \\
NGC~4258-8    & GO-16743\tnote{e}   & 15.0   & 883  & 0.227 & $25.352^{+0.022}_{-0.028}$ & $25.302 \pm 0.041$ & $25.312^{+0.022}_{-0.028}$ & $25.331 \pm 0.041$ \\
NGC~4258-9    & GO-16743\tnote{e}   & 12.5 & 438  & 0.147 & $25.325^{+0.030}_{-0.035}$ & $25.320 \pm 0.041$ & $25.286^{+0.030}_{-0.035}$ & $25.330 \pm 0.041$ \\
NGC~4258-10   & GO-16743\tnote{e}   & 15.0   & 1020 & 0.153 & $25.308^{+0.016}_{-0.016}$ & $25.268 \pm 0.040$ & $25.267^{+0.016}_{-0.016}$ & $25.382 \pm 0.040$ \\
\bottomrule
\end{tabular}}
\begin{tablenotes}
\small
\item[a] \cite{madore_trgb_2002}; \textsuperscript{b}\cite{greenhill_accurate_2004}; \textsuperscript{c}\cite{riess_masers_2020}; \textsuperscript{d}\cite{anderson_towards_2021}; \textsuperscript{e}\cite{hoyt_high-accuracy_2021}
\item[f] Dereddened for Milky Way foreground (\(A_{814}=0.03\); \citealt{schlafly_measuring_2011}) and per-field internal extinction (0.009--0.019~mag; \citealt{menard_measuring_2010}, as tabulated by \citealt{li_standardized_2023}).
\item[g] Same as in footnote~f, and additionally standardised to a fiducial contrast ratio \(R=4\) via the tip--contrast relation (TCR; \citealt{li_standardized_2023}), where \(R\) is the ratio of star counts in 0.5~mag below the tip to 0.5~mag above it. The ratios \(R\) and \(r\) are distinct and run in the opposite sense: a large \(R\) (which can go to infinity) and a small \(r\) correspond to a well-defined tip. This work instead absorbs the field-to-field variation in the tip into the galaxy-level scatter \(\tau\), so the corrected ``this work'' tips are dereddened but not TCR-standardised.
\end{tablenotes}
\end{threeparttable}
\end{table*}

To test how our inference procedure performs with real data, we analyse \textit{HST} imaging of 11 NGC~4258 fields, as listed in Table~\ref{tab:ngc4258_fields}. This imaging was reduced photometrically using DOLPHOT \citep{dolphin_wfpc2_2000, dolphin_dolphot_2016}, following the same pipeline and quality cuts as \cite{anand_comparing_2022}. We take the per-star photometry for these fields from the Comparative Analysis of TRGBs (CATs) sample of \cite{li_standardized_2023}, publicly available at \url{https://github.com/JiaxiWu1018/CATS-H0} and produced as part of the broader CATs project \citep{wu_comparative_2023, scolnic_cats_2023}, which applies a spatial clip and colour-band selection to the photometry to isolate RGB stars. We adopt their cleaned per-star magnitudes and uncertainties as inputs to the framework developed here, rather than their own tip-detection algorithm. 

Flux is expressed throughout this section in physical units of \microjansky{}, via \(f/f_0 \equiv 10^{-2m/5}\), where \(m\) is the VEGAMAG (F814W) magnitude reported by DOLPHOT and \(f_0\) is the ACS/WFC F814W Vega reference flux for the field's specific \textit{HST} imaging epoch (obtained from STScI's photometric calibration tables; \citealt{lim_acstools_2020}); both the flux and the magnitude are therefore physically calibrated quantities. We work in flux throughout the model, but quote inferred tips and their variation in magnitudes for comparison with published values.

\cite{li_standardized_2023} report TRGB values for each field before and after applying their extinction and tip--contrast relation (TCR) corrections. The latter is an empirical correction for the observed dependence of the measured tip on contrast ratio. \cite{anderson_small-amplitude_2024} show that the edge-detection response weighting used in the CATs algorithm can itself create the need for a post-edge-detection TCR correction. By contrast, our Bayesian model fits the underlying LF directly: there is no smoothing of the data, no weighting, and no derivative-peak search that could introduce bias. 

For each field, we examine the \((\hat{f}_i,\hat{\sigma}_i)\) distribution in the F814W filter and remove stars that have a \((\hat{f},\hat{\sigma})\) coordinate that sits far from the distinct locus. As described in Section~\ref{sec:Application to catalogue-level data}, we fit the parametric form \(\hat{\sigma} =\sqrt{\hat{\sigma}_0^2 + \hat{C}\hat{f}}\) to the observed \((\hat{f}_i, \hat{\sigma}_i)\) distribution in order to calculate \(\Nbar\). 

Fluxes far (\(\gg \sigma\)) from \(\fTRGB\) carry little information about the position of the discontinuity, so we are free to impose our own flux cut and reduce the number of latent fluxes the model must infer. The caveats and validity of this choice are discussed further in Section~\ref{sec:Discussion} and Appendix~\ref{app:rho_sweep}. The CATs catalogue already applies its own selection cut, which we measure to be at \(\mathrm{SNR}=4\). We can impose our own more conservative cut, provided that it is chosen sufficiently far above the catalogue's completeness limit so that no star that should enter our sample is missing from theirs. For each field, we adopt the cut at the midpoint of the range over which the inferred tip value is stable, determined as described in Appendix~\ref{app:rho_sweep}. 

The \(\Nbar\) integral we use assumes a cut at a chosen flux value, which differs from a strict SNR cut, which would leave a fuzzy edge to the data due to the scatter around the \((\hat{f}_i, \hat{\sigma}_i)\) locus. In the absence of scatter about the locus, our cut is valid for \(\fcut/\hat\sigma(\fcut)\equiv \rhostar>4\). In practice, the retained stars scatter to uncertainties above \(\hat\sigma(\fcut)\), which tightens the constraint to \(\rhostar>5.4\) for all 11 NGC~4258 fields. This is discussed further in Appendix~\ref{app:rho_sweep}, where we also verify that the inferred tip is insensitive to \(\rhostar\) over a wide range. 

We begin with a detailed analysis of two extreme fields: NGC~4258-10 (Section~\ref{sec:field 10}), which has high \(N\) and a visually clear TRGB, so it is close to an ideal case; and NGC~4258-5 (Section~\ref{sec:field 5}), which gives two peaks in the edge-detection algorithm of \cite{li_standardized_2023} and so may not yield good results. We then analyse the remaining nine fields and combine all 11 into a single galaxy-level measurement for NGC~4258 (Section~\ref{sec:TRGB magnitude}).

\subsection{NGC~4258-10}
\label{sec:field 10}

The \((\hat{f}_i,\hat{\sigma}_i)\) distribution for Field~10 shows a distinct locus that holds 99.3\% of the fluxes, leaving 3690 \((\hat{f},\hat{\sigma})\) pairs. We fit the parametric form of Section~\ref{sec:Application to catalogue-level data} to these pairs and use the \(\hat{\sigma}_0=0.0028\)~\microjansky{} and \(\hat{C}=0.000057\)~\microjansky{} values in the computation of \(\Nbar\) in the Stan model. We apply the cut-choice procedure, which gives \(\rhostar=15\), and \(\fcut=0.048\)~\microjansky{}, defined by Equation~\eqref{eqn:fcut}, leaving 1020 stars in the final sample. 

\begin{figure}
    \centering
    \includegraphics[width=\columnwidth]{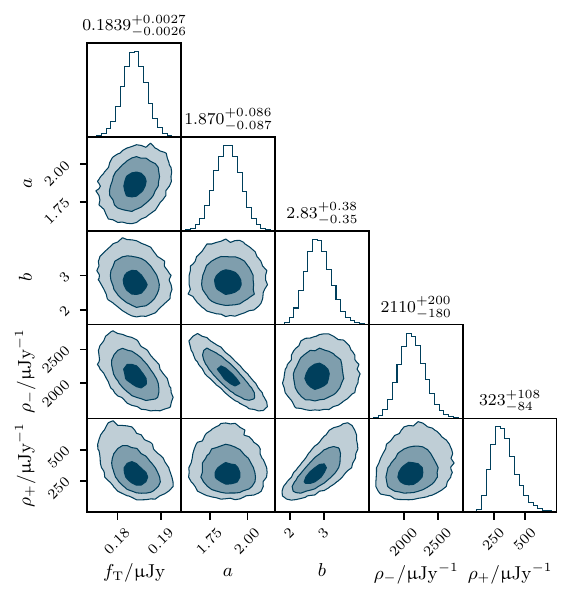}
    \caption{Corner plot showing the marginal posteriors of the five population-level parameters \(\fTRGB\), \(a\), \(b\), \(\rhobelow\), and \(\rhoabove\) for NGC~4258-10. Filled contours enclose 39.3\%, 86.5\%, and 98.9\% of the posterior probability in each panel. The value above each marginal is its posterior median with 16th and 84th percentiles.}
    \label{fig:corner_NGC4258-10}
\end{figure}

\begin{figure}
    \centering
    \includegraphics[width=\columnwidth]{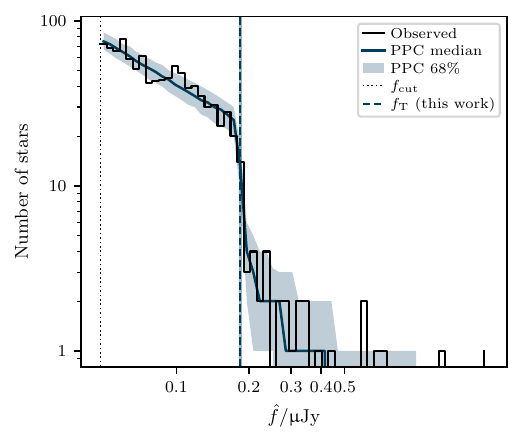}
    \caption{Observed flux distribution for NGC~4258-10 (step histogram), compared with the posterior predictive distribution from 1000 replicates. The solid line and shaded band show the median and 16th to 84th percentile range of the replicate counts in each bin. The dashed vertical line and its shaded band show the inferred \(\fTRGB\) median and 16th to 84th percentile range. The flux cut, \(\fcut\), is plotted as a dotted vertical line.}
    \label{fig:PPC_NGC4258-10}
\end{figure}

Sampling from the posterior gives a median and 16th to 84th percentiles of \(\fTRGB = 0.1839^{+0.0027}_{-0.0026}\)~\microjansky{}, corresponding to \(\mTRGB=25.308 \pm 0.016\)~mag. After correcting for the internal (\(A_{814}=0.011\); \citealt{menard_measuring_2010}) and Milky Way (\(A_{814}=0.03\); \citealt{schlafly_measuring_2011}) extinctions tabulated in \cite{li_standardized_2023}, we find a magnitude of \(25.267 \pm 0.016\). As shown in Figure~\ref{fig:corner_NGC4258-10}, the posterior found for NGC~4258-10's population-level parameters is close to a multivariate normal, apart from some expected correlation between the shape parameters \(a\) (for the RGB) and \(b\) (for the AGB), and their corresponding densities \(\rhobelow\) and \(\rhoabove\). In Figure~\ref{fig:PPC_NGC4258-10}, we compare the observed flux distribution with the posterior predictive distribution \citep{rubin_bayesianly_1984, gelman_posterior_1996}. For each of the 1000 replicates, we draw from the population-level posterior and follow the full generative model (Section~\ref{sec:Generative Model}), producing a set of measured fluxes. The 68\% band therefore shows the catalogues that the model predicts, not only the shape of the LF. The posterior predictive \(p\)-values for the median and interquartile range of the flux distribution are 0.268 and 0.480, suggesting a good fit to the data. 

\begin{figure}
    \centering
    \includegraphics[width=\columnwidth]{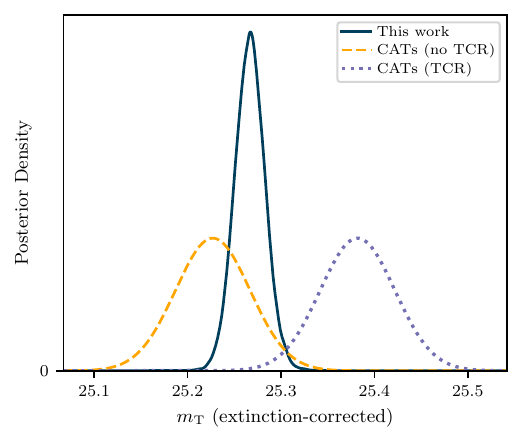}
    \caption{Comparison of this work's NGC~4258-10 \(\mTRGB\) posterior distribution to the assumed-normal posteriors of CATs, with and without their TCR correction. All are extinction-corrected, so the offset between the two CATs curves is the tip--contrast standardisation alone.}
    \label{fig:m_T_vs_cats_NGC4258-10}
\end{figure}

\cite{li_standardized_2023} report TRGB values for the NGC~4258-10 field in F814W of \(25.268\pm 0.040\) before applying their extinction and TCR corrections, and \(25.382\pm 0.040\) after. As shown in Figure~\ref{fig:m_T_vs_cats_NGC4258-10}, where all values are corrected for extinction, our posterior for the tip magnitude sits between the two (assumed normal) CATs distributions, although closer to their value before TCR standardisation.

\subsection{NGC~4258-5}
\label{sec:field 5}

One of the advantages of using a fully Bayesian pipeline is that all the information is held in the posterior: if a dataset carries genuine ambiguity about the location of the TRGB, the posterior will reflect this --- with the distribution wide, skew, or (in extreme cases) multimodal --- rather than collapsing to a single point estimate. As \cite{li_standardized_2023} find two TRGB values for Field~5, the field will be a good test of how the Bayesian model performs in less ideal situations.

We keep stars that follow a well-defined locus in \((\hat{f},\hat{\sigma})\) space, leaving 5230 stars (94.6\%). We fit the same parametric form to the remaining \((\hat{f},\hat{\sigma})\) pairs, giving \(\hat{\sigma}_0=0.0065\)~\microjansky{} and \(\hat{C}=0.00026\)~\microjansky{}. We then cut at \(\fcut=0.089\)~\microjansky{}, corresponding to a \(\rhostar = 11\) cut. This leaves 1381 stars in the final sample. We find \(\fTRGB =0.1538^{+0.0062}_{-0.0088}\)~\microjansky{}, corresponding to \(\mTRGB=25.501^{+0.064}_{-0.043}\)~mag. After correcting for the internal (\(A_{814}=0.016\); \citealt{menard_measuring_2010}) and Milky Way (\(A_{814}=0.03\); \citealt{schlafly_measuring_2011}) extinctions, we find a magnitude of \(25.455^{+0.064}_{-0.043}\).

\begin{figure}
    \centering
    \includegraphics[width=\columnwidth]{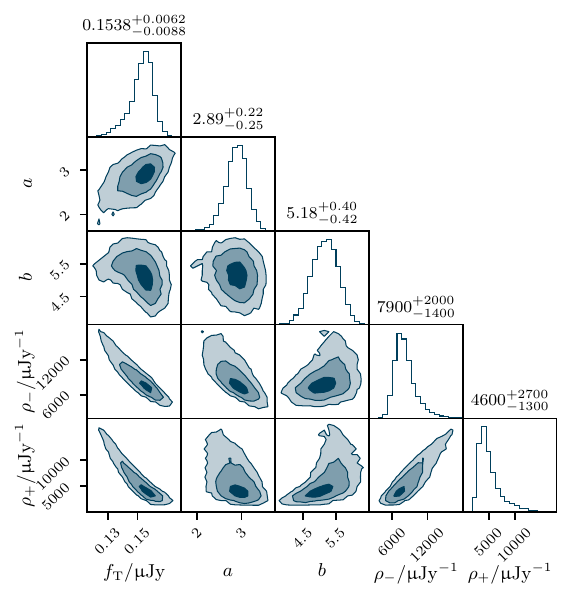}
    \caption{Corner plot showing the marginal posteriors of the five population-level parameters \(\fTRGB\), \(a\), \(b\), \(\rhobelow\), and \(\rhoabove\) for NGC~4258-5. Filled contours enclose 39.3\%, 86.5\%, and 98.9\% of the posterior probability in each panel. The value above each marginal is its posterior median with 16th and 84th percentiles.}
    \label{fig:corner_NGC4258-5}
\end{figure}

\begin{figure}
    \centering
    \includegraphics[width=\columnwidth]{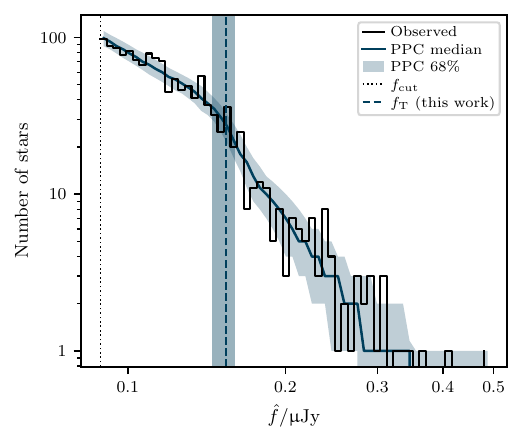}
    \caption{Observed flux distribution for NGC~4258-5 (step histogram), compared with the posterior predictive distribution from 1000 replicates. The solid line and shaded band show the median and 16th to 84th percentile range of the replicate counts in each bin. The dashed vertical line and its shaded band show the inferred \(\fTRGB\) median and 16th to 84th percentile range. The flux cut, \(\fcut\), is plotted as a dotted vertical line.}
    \label{fig:PPC_NGC4258-5}
\end{figure}

As shown in Figure~\ref{fig:corner_NGC4258-5}, the posterior is not a multivariate normal, and the marginal posterior of \(\fTRGB\) is skewed leftwards towards lower flux. The posterior predictive distribution for Field~5 is shown in Figure~\ref{fig:PPC_NGC4258-5}. With 1000 replicates, we find posterior predictive \(p\)-values for the median and interquartile range of the flux distribution of 0.44 and 0.46, neither indicating a discrepancy. Comparing the flux distributions of Fields~5 and 10, we see that Field~5 does not display as clean a discontinuity in star count as Field~10 does. Instead, the model locates the tip through the change in slope between the RGB and AGB. 

\begin{figure*}
    \centering
    \includegraphics[width=\textwidth]{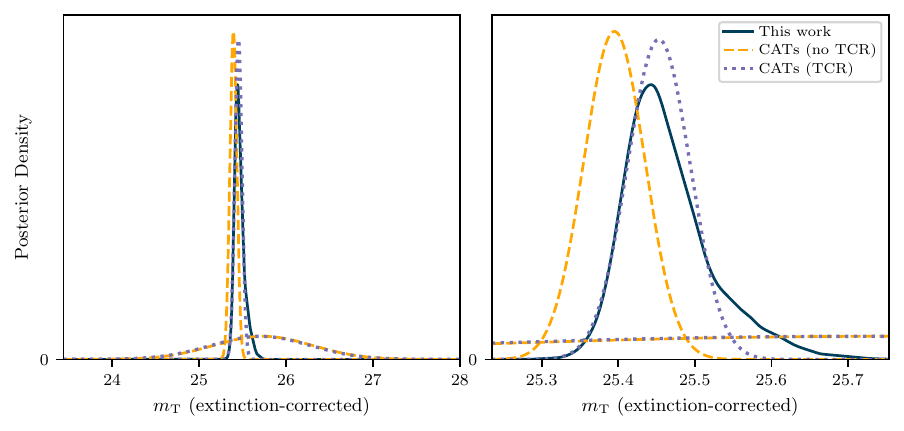}
    \caption{Comparison of this work's NGC~4258-5 \(\mTRGB\) posterior distribution to the assumed-normal posteriors of CATs, with and without their TCR correction. CATs report two detections for this field, and both are plotted; the broad pair is their second detection. The right panel is an enlarged view of the left panel. All are extinction-corrected, so the offset between each pair of CATs curves is the tip--contrast standardisation alone.}
    \label{fig:m_T_vs_cats_NGC4258-5}
\end{figure*}

\cite{li_standardized_2023} find two separate peaks in their algorithm for
Field~5, at \(25.441 \pm 0.041\) and \(25.781 \pm 0.567\)~mag before TCR and
extinction correction, and \(25.453 \pm 0.041\) and \(25.707 \pm 0.567\)~mag
after. Figure~\ref{fig:m_T_vs_cats_NGC4258-5} shows the two (assumed normal and all corrected for extinction) pairs of posteriors from the \cite{li_standardized_2023} algorithm, and the marginal posterior of \(\mTRGB\) from our model. The direction of the skew of the posterior is informative: the heavy right tail towards fainter magnitudes suggests that there is some truth to the two-tip detection of the CATs algorithm. The ambiguity between the two peaks is expressed as posterior skew, rather than a choice between two point estimates.

\subsection{Combining the fields}
\label{sec:TRGB magnitude}

Having examined one benign and one difficult field in detail, we analyse the other fields in \cite{li_standardized_2023} similarly and report our findings in Table~\ref{tab:ngc4258_fields}, along with the \(\rhostar\) chosen for each field as determined by the protocol of Appendix~\ref{app:rho_sweep}, and the inferred \(r\equiv\rhoabove/\rhobelow\). The extreme skew of the \(\fTRGB\) posterior for Field~2 is an artefact of a second smaller peak found fainter than the main peak. 

The TRGB magnitudes found for the 11 fields span a range of 0.229~mag, with a sample standard deviation of 0.073~mag, considerably larger than the quoted precision of some of the individual fields. This implies that there is real field-to-field variation on top of measurement noise, and combining the fields using inverse-variance weighting (which assumes that the fields measure an identical tip) would both understate the combined uncertainty and overweight the most precise fields.

We therefore treat the extinction-corrected tips of Table~\ref{tab:ngc4258_fields} as noisy realisations of a single galaxy-level TRGB, \(\mgalaxy\), with an intrinsic field-to-field scatter, \(\tau\), that we infer jointly with \(\mgalaxy\). This allows the data themselves to determine whether such scatter is present: an inferred \(\tau\simeq0\) would suggest that there is no scatter beyond measurement uncertainty.

We assume a normal prior on the \(j^{\mathrm{th}}\) field's true TRGB magnitude, with mean \(\mgalaxy\) and variance \(\tau^2\),
\begin{equation}
    \mjfield \sim \mathcal{N}(\mgalaxy,\tau^2).
\end{equation}
The observed stellar catalogue, \(D_j\), of field \(j\) is then generated from its own tip, \(\mjfield\), as described in Section~\ref{sec:Generative Model}. Using Bayes' theorem, the joint posterior on the galaxy-level parameters and field-level tip magnitudes is
\begin{multline}
    P(\msetjfield, \mgalaxy, \tau \mid \Dsetj) \\
    \propto \pi (\mgalaxy, \tau) \prod^{11}_{j=1} P(D_j\mid\mjfield) \,\mathcal{N}(\mjfield;\mgalaxy, \tau^2),
\end{multline}
where \(\pi (\mgalaxy, \tau)\) is the prior on the galaxy-level parameters. We marginalise over each field's tip magnitude, giving
\begin{multline}
    P(\mgalaxy, \tau \mid \Dsetj) \\
    \propto \pi (\mgalaxy, \tau) \prod^{11}_{j=1}\int \mathrm{d} \mjfield \,   P(D_j\mid\mjfield) \,\mathcal{N}(\mjfield;\mgalaxy, \tau^2).
\end{multline}
Because the field-level prior on \(\mjfield\) is flat (due to the uniform prior on \(\log{\fTRGB}\) from Section~\ref{sec:sampling}), the posterior simplifies to
\begin{multline}
    P(\mgalaxy, \tau \mid \Dsetj) \\
    \propto \pi (\mgalaxy, \tau) \prod^{11}_{j=1}\int \mathrm{d} \mjfield \,   P(\mjfield \mid D_j) \,\mathcal{N}(\mjfield;\mgalaxy, \tau^2).
\end{multline} 

With \(K\) samples from the posterior for each field, \(\{\mjfield^{(k)}\}_{k=1}^{K}\), we can use Monte Carlo integration to approximate the joint posterior as
\begin{equation}
P(\mgalaxy, \tau \mid \Dsetj)
    \propto \pi(\mgalaxy)\,\pi(\tau)\,\prod^{11}_{j=1}\frac{1}{K}\sum^K_{k=1} \mathcal{N}(\mjfield^{(k)};\mgalaxy,\tau^2).
    \label{eqn:hier_posterior}
\end{equation}
Using the draws rather than approximating each field-level posterior as normal allows the galaxy-level inference to account for their (often extreme) skewness. 

We adopt a flat prior on \(\mgalaxy\) and a weakly informative half-Cauchy prior, \(\tau \sim \mathrm{Half\text{-}Cauchy}(0, 0.1)\) \citep{gelman_prior_2006}, and sample the posterior using Stan. We also repeat the inference with half-Cauchy scales of 0.05 and 0.5, a half-normal, a uniform, and a log-uniform prior on \(\tau\), and find \(\mgalaxy\) changes by \(<0.001\)~mag in all cases, and \(\tau\) by \(<0.01\)~mag.

We find \(\mgalaxy=25.324 \pm 0.023\)~mag, and \(\tau=0.067^{+0.023}_{-0.017}\)~mag. A non-zero \(\tau\) confirms that there is an intrinsic field-to-field scatter above measurement uncertainty, which must be included in our error budget: a naive inverse-variance combination gives \(25.287\pm 0.009\)~mag, which is 0.037~mag brighter and has an unjustifiably small uncertainty. The flux-cut policy carries an uncertainty of 0.012~mag, as discussed in Appendix~\ref{app:rho_sweep}, which is added in quadrature to the posterior width to give a total uncertainty of \(\pm 0.026\)~mag. Subtracting the megamaser distance modulus to NGC~4258 of \(29.397 \pm 0.032\)~mag \citep{reid_improved_2019}, we find an absolute magnitude of the TRGB in the F814W filter of 
\begin{equation*}
    \MTRGBHSTband = -4.073 \pm 0.026\;\text{(tip)} \pm 0.032\;\text{(dist)}.
\end{equation*}
Extinction corrections are treated as exact to match the CATs treatment against which we compare. The error budget is tabulated in Table~\ref{tab:error_budget}.

\begin{table}
\centering
\caption{Error budget for the galaxy-level F814W TRGB magnitude of NGC~4258. Extinction corrections are treated as exact and are not included. Terms are combined in quadrature.}
\label{tab:error_budget}
{\renewcommand{\arraystretch}{1.2}
\begin{tabular}{llr}
\toprule
Source & Term & (mag) \\
\midrule
Statistical & Hierarchical posterior on \(\mgalaxy\)        & 0.023 \\
Systematic  & Flux-cut policy                               & 0.012 \\
\cmidrule(l){2-3}
\multicolumn{2}{l}{Tip}   & \textbf{0.026} \\
\midrule
Distance    & NGC~4258 megamaser \citep{reid_improved_2019}  & 0.032 \\
\bottomrule
\end{tabular}}
\end{table}

A set of recent calibrations of \(\MTRGBHSTband\) is listed in Table~\ref{tab:f814W_M_TRGB_values}. Our value falls on the bright side of the current estimates, although well within the combined uncertainties. The three unstandardised measurements span 0.044~mag across two independent anchors (NGC~4258 megamaser, LMC eclipsing binaries) and three independent pipelines. However, the two NGC~4258 values are derived in part using the same imaging (GO-9477 and GO-10399). The two most discrepant fields in our analysis, Fields~5 and 6, are on the fainter side of the final estimate and do not drive our value brighter. With these fields excluded, we find \(\mgalaxy=25.293 \pm 0.017\)~mag, and \(\tau=0.038_{-0.014}^{+0.019}\)~mag, a shift of 0.031~mag, larger than the statistical uncertainty, and \(\MTRGBHSTband = -4.104 \pm 0.021\;\text{(tip)} \pm 0.032\;\text{(dist)}\), further from published calibrations.

\begin{table*}
\centering
\caption{Recent absolute magnitude calibrations of the TRGB in the \textit{HST} F814W filter. The entries differ in standardisation as well as in anchor, which accounts for much of the spread. See \cite{freedman_measurements_2021} for a compilation of earlier \(I\)-band determinations.}
\label{tab:f814W_M_TRGB_values}
\begin{threeparttable}
{\renewcommand{\arraystretch}{1.3}
\begin{tabular}{lccl}
\toprule
Reference & Anchor & \(\MTRGBHSTband\) (mag) & Standardisation \\
\midrule
This work                            & NGC~4258 & \(-4.073 \pm 0.026 \pm 0.032\)\tnote{a} & none \\
\cite{jang_carnegie-chicago_2021}     & NGC~4258 & \(-4.050 \pm 0.028 \pm 0.048\)          & none\tnote{b} \\
\cite{li_standardized_2023}           & NGC~4258 & \(-4.036 \pm 0.035\)                    & \(R=4\) \\
\cite{freedman_calibration_2020}      & LMC      & \(-4.054 \pm 0.022 \pm 0.039\)          & \((V-I)=1.6\) \\
\cite{udalski_ultimate_2025}      & LMC      & \(-4.029 \pm 0.006 \pm 0.033\)\tnote{c} & none \\
\cite{freedman_measurements_2021}     & MW, NGC~4258, LMC, SMC      & \(-4.049 \pm 0.015 \pm 0.035\)          & \((V-I)=1.6\) \\
\bottomrule
\end{tabular}}
\begin{tablenotes}
\small
\item[a] Uncertainties are (tip) and (distance), respectively; other entries are (stat) and (sys).
\item[b] Measured from a spatially selected halo sample (semimajor axis \(>14'\)) rather than standardised.
\item[c] Published as \(M_I = -4.022 \pm 0.006 \pm 0.033\) and transformed to F814W using
\(M_\mathrm{F814W} = M_I - 0.0068\)~mag at \((V-I)=1.6\) \citep{freedman_calibration_2020}.
\end{tablenotes}
\end{threeparttable}
\end{table*}

\cite{jang_carnegie-chicago_2021} use Fields~1, 2, 3, 4~G1, and 4~G2 in their analysis of NGC~4258. With this field selection, we find \(\MTRGBHSTband = -4.072 \pm 0.026\;\text{(tip)} \pm 0.032\;\text{(dist)}\), which is in agreement with their value. We find that \(\tau\) is consistent with 0 with this field selection. 

\section{Discussion}
\label{sec:Discussion}

The absolute magnitude of the TRGB we find, \(\MTRGBHSTband = -4.073 \pm 0.026\;\text{(tip)} \pm 0.032\;\text{(dist)}\), lies at the bright end of literature values, as shown in Table~\ref{tab:f814W_M_TRGB_values}, especially when compared directly to \cite{li_standardized_2023}, whose data we use. That there is a difference is unsurprising, as we are measuring two different estimands: the CATs algorithm measures, at its core, the peak in the gradient of a smoothed, binned LF, whereas our pipeline treats fluxes as individual draws from an underlying population with a true LF. Even with this difference in methodology, there is a general agreement with the CATs TRGBs: the field-level tips agree within \(2\sigma\) before tip--contrast standardisation, the worst case being \(1.8\sigma\) (Table~\ref{tab:ngc4258_fields}). The gap in the published values arises after standardisation, which we do not apply. 

Whilst it might seem appealing to use this new calibration of \(\MTRGBHSTband\) to obtain distances to SNe Ia host galaxies to infer a value of \(H_0\), the TRGBs of those hosts must also be analysed using this method before a self-consistent \(H_0\) can be reported. This work serves to introduce a Bayesian method for inferring the TRGB, with future work focused on SNe Ia host galaxies. 

The method produces a marginal posterior for each parameter in the model, whose shape can be used as additional information. Whilst the \(\fTRGB\) marginal posteriors of many of the well-defined tips are in an asymptotically normal regime, it is evident from some of the fields (\textit{e.g.}\@ Field~2) that assuming a normal shape is incorrect and hides multimodality and skew. Although this extra information comes at a higher computational cost than edge-detection algorithms --- the CATs algorithm is far faster --- the use of HMC means that each fit takes \(\sim\!10\) minutes on a laptop, rather than requiring any more advanced hardware. 

Using a galaxy-level hierarchical model, we quantify the field-to-field scatter for NGC~4258, an open problem identified by \cite{valentino_cosmoverse_2025}. We do not propose that the value of \(\tau = 0.067\)~mag found for NGC~4258 holds for all galaxies; it is a measurement for NGC~4258 alone. An important consequence of this is that any single-field analysis of NGC~4258 cannot account for field-to-field variation in the tip across the galaxy, which, for the five most precise fields, exceeds the quoted per-field uncertainty by a factor of two or more. Future TRGB analyses should include this uncertainty; the hierarchical method used here measures and propagates the variation directly. 

Whilst Section~\ref{sec:error forecast} forecasts per-field uncertainty, Section~\ref{sec:Real data} has shown that the fields within a galaxy might not share a TRGB value. Therefore, the galaxy-level uncertainty contains a \(\tau/\sqrt{J}\) term that will not reduce with increased depth or density of single fields. For sufficiently precise fields, as shown above, the field-to-field scatter provides a hard lower bound to the uncertainty, and only increasing the number of fields helps to reduce the uncertainty. Where fields are not more precise than the field-to-field scatter, the most important choice a survey can make is identified in Section~\ref{sec:error forecast}: reducing \(r\), the ratio of the stellar density brighter than the tip to that fainter than it, which a survey can control through field placement. 

As listed in Table~\ref{tab:ngc4258_fields}, Fields~5 and 6 have the largest and second-largest \(r\) values among the NGC~4258 fields, at \(r = 0.593\) and \(r = 0.519\), respectively. Section~\ref{sec:Sensitivity to AGB Contamination} identifies \(r\gtrsim 0.5\) as a region where the tip becomes poorly defined, and the \(\fTRGB\) posteriors are found to be multimodal. Whilst neither field's \(\fTRGB\) posterior is multimodal, both fields show poor tip identifiability and a narrow stable cut range. Field~9 has the lowest \(r\) (0.147), with Fields~10, 8, and 7 the next lowest. These fields hold four out of the five most precise (ranked by the width of the 68\% interval) \(\fTRGB\) posteriors. Field~1 has the third-most precise posterior, but has the third-largest \(r\) (0.472), suggesting that a high \(r\) does not guarantee a wide posterior distribution. Indeed, in the limit of large \(N\), we expect to recover a tip to arbitrarily high precision for any \(r\). Further, a low \(r\) is a useful but not sufficient predictor of a precise tip, as evidenced by Field~4~G1, which has the fifth-lowest \(r\) of 0.336 but the second-widest posterior. The number of stars observed, \(N\), is not a good predictor of precision either: Field~5 has the most stars and ranks eighth in posterior precision, whilst Fields~2 and 9 have the same \(N=438\) yet rank eleventh and fifth, respectively.

The model only describes the one-dimensional LF, and several of its limitations follow from this choice. Colour information is discarded, and we measure each field's own tip without standardisation across the galaxy. We also model the tip as a discontinuity, meaning that any intrinsic smearing from age and metallicity spread within a field is absorbed into its tip measurement, rather than kept separate. Separately, Appendix~\ref{app:rho_sweep} finds curvature in the RGB LF away from the power law that the model assumes, whose origin we do not identify. A two-dimensional treatment could address both, since colour carries information about the population composition that is inaccessible in one dimension, whilst keeping the hierarchical model form. 

\section{Conclusions}
\label{sec:conclusion}
We present a Bayesian forward model to infer the flux of the TRGB, treating the stellar catalogue as draws from an inhomogeneous Poisson point process, where selection is modelled rather than conditioned upon, and per-star uncertainties are taken into account. We validate the model by fixed-value calibration. 

We also use a simplified model of the LF to derive closed-form expressions for the uncertainty in the TRGB flux. With the uncertainty described by Equation~\eqref{eqn:bvm_step}, we find that at a fixed total exposure time, a wide, shallow survey is preferable to a narrow, deep one --- observing more stars is more important than reducing photometric uncertainty. Future surveys should also be designed to reduce \(r\) by imaging fields far from the disc centre with lower AGB contamination, whilst ensuring that there is a sufficient surface density of RGB stars to allow for TRGB identification. 

Applying our method to \textit{HST} data from 11 fields covering NGC~4258, we find that the field-level TRGBs are more widely spread than expected given their uncertainties, so we combine them using a galaxy-level hierarchical model that infers the galaxy TRGB magnitude jointly with an intrinsic field-to-field scatter. Anchored on the NGC~4258 megamaser distance, we measure a calibration of the TRGB magnitude in the F814W filter of \(\MTRGBHSTband = -4.073 \pm 0.026\;\text{(tip)} \pm 0.032\;\text{(dist)}\), and an intrinsic field-to-field scatter within that galaxy of \(\tau = 0.067^{+0.023}_{-0.017}\)~mag. No single-field analysis can see this scatter, which exceeds the per-field uncertainty by a factor of two or more for the five most precise fields. The final uncertainty quoted for a galaxy should therefore include a \(\tau/\sqrt{J}\) floor over \(J\) fields, which no increase in depth will reduce.

The obvious next step in this project is to apply our methodology to SNe Ia host galaxies to infer \(H_0\). An important part of this will be to measure the scatter, \(\tau\), for other galaxies to assess whether it is specific to NGC~4258 or a more general property. One possibility for reducing the impact of this scatter is to exploit the fact that observed CMDs exhibit structure perpendicular to the RGB, which should provide information on \textit{e.g.}\@ stellar metallicity. Therefore, we intend to extend our approach to go beyond the one-dimensional case (\textit{i.e.}\@ just flux/magnitude counts) to the two-dimensional case with colour information as well.

\section*{Acknowledgements}

JDG is supported by STFC grant ST/Y509231/1.
We thank Siyang Li for providing previously unpublished data, and Suhail Dhawan, Alan Heavens, Andrew Jaffe, and Dan Scolnic for their valuable comments.

\bibliographystyle{mnras}
\bibliography{references}

\appendix
\renewcommand{\theHequation}{\thesection.\arabic{equation}}
\renewcommand{\theHtable}{\thesection.\arabic{table}}
\renewcommand{\theHfigure}{\thesection.\arabic{figure}}

\section{Convergence of the Pareto Distribution}
\label{app:pareto distribution}

We wish to show that, in the no-AGB, zero-noise limit of Section~\ref{sec:no noise no agb}, the Pareto distribution of Equation~\eqref{eqn: no noise no agb pareto posterior} converges to the exponential distribution of Equation~\eqref{eqn:no noise no AGB exponential} for large \(N\), and hence large \(\rhobelow\). Making the substitution of Equation~\eqref{eqn:fhat substitution}, we find
\begin{equation}
    P(\fTRGB \mid \hat{f}_{1:N}, N) \simeq
    \Theta(\fTRGB-\fThat)\,\rhobelow\left(\frac{\fTRGB}{\fThat}\right)^{-(\rhobelow\fThat+1)}.
\end{equation}
Using 
\begin{equation}
    \lim_{n\rightarrow\infty}\left(1+\frac{x}{n}\right)^{-n}=\exp(-x),
\end{equation}
setting \(x=\rhobelow(\fTRGB-\fThat)\) and \(n=\rhobelow\fThat\), and dropping the \(+1\) that is negligible for large \(\rhobelow\),
\begin{align}
    P(\fTRGB \mid \hat{f}_{1:N}, N)&\simeq  \Theta(\fTRGB-\fThat)\,\rhobelow\left(1+\frac{\fTRGB-\fThat}{\fThat}\right)^{-\rhobelow\fThat}\notag\\
    &\propto\Theta\!\left(\frac{x}{\rhobelow}\right)\left(1+\frac{x}{\rhobelow\fThat}\right)^{-\rhobelow\fThat}\notag\\
    &\simeq\Theta(x) \, \exp(-x),
\end{align}
leaving
\begin{equation}
    P(\fTRGB \mid \hat{f}_{1:N}, N) \propto \Theta(\fTRGB-\fThat)\, \exp(-\rhobelow\fTRGB),
\end{equation}
which is a shifted exponential distribution with parameter \(\rhobelow\).

\section{Fisher Information of an Inhomogeneous Poisson Point Process}
\label{app:Fisher Information}

For a parametric model with log-likelihood \(\ell(X, \theta)\), where \(X\) denotes the data and \(\theta\) the model parameters, the Fisher information matrix has entries
\begin{equation}
    \mathcal{I}_{lm}(\theta) =
    -\mathrm{E}\!\left[\frac{\partial^2 \ell}{\partial \theta_l\,
    \partial \theta_m}\right],
\end{equation}
where the expectation is taken with respect to the data under the model. For an inhomogeneous PPP with intensity \(\lambda(s,\theta)\) on a region \(R\), where \(s\) is a position in that region, with a \(d\)-dimensional parameter vector \(\theta\), this reduces to \citep[Section~2.1, pp.~54, 59]{kutoyants_introduction_2023}
\begin{equation}
    \mathcal{I}_{lm}(\theta) = \int_R \mathrm{d} s\, \frac{1}{\lambda(s,\theta)}\,\frac{\partial \lambda}{\partial \theta_l}\,\frac{\partial \lambda}{\partial \theta_m}\,,\qquad l,m = 1,\ldots,d\,.
    \label{eqn:fisher-information-PPP}
\end{equation}

We apply this in Section~\ref{sec:error forecast}, with \(s=\hat{f}\), \(R= \{\hat{f}>\fcut\}\), and intensity \(\lambda(\hat{f},\theta) = \int_{0}^{\infty} \mathrm{d}f\, \psi(f,\theta)\, P(\hat{f}\mid f)\).

\section{Sensitivity to the Flux Cut}
\label{app:rho_sweep}

As discussed in Section~\ref{sec:Real data}, we apply a flux cut to the data from the CATs catalogue, derived by fitting the observed locus in \((\hat{f},\hat{\sigma})\) space with the parametric form \(\hat{\sigma} =\sqrt{\hat{\sigma}_0^2 + \hat{C}\hat{f}}\). We then derive \(\fcut\) using Equation~\eqref{eqn:fcut}. 

We expect the model to fail at extreme \(\fcut\) values. Trivially, if \(\fcut\) is selected such that \(\fcut>\fTRGB\), the model will not see the TRGB and will not infer the correct tip. In practice, we want \(\fTRGB-\fcut\gg \hat\sigma(\fTRGB)\), \textit{i.e.}\@ we want the cut to be sufficiently far below the TRGB for there to be no chance that any stars with \(f>\fTRGB\) have scattered below the cut. This defines the maximum acceptable value for \(\fcut\). We also need enough RGB stars with fluxes between the cut and \(\fTRGB\) for the tip to be identifiable. 

Even if the uncertainty follows the \((\hat{f},\hat{\sigma})\) locus perfectly with no scatter, choosing an \(\fcut/\hat{\sigma}(\fcut)< \mathrm{SNR}\) for some SNR-limited catalogue guarantees model misspecification: whilst the model expects a cut at \(\fcut\), the data are bounded by the SNR cut instead. 

In real data, this lower limit is tightened by scatter around this locus. An SNR cut at \(\rhostar\) defines a straight line through the origin in \((\hat{f},\hat{\sigma})\) space with gradient \(1/\rhostar\), with data above the line removed from the catalogue, and data below the line kept. A constant flux cut at \(\fcut\) defines a vertical line in this space at \(\hat{f}=\fcut\), keeping all data to the right of it. With scatter around the \(\hat{\sigma} =\sqrt{\hat{\sigma}_0^2 + \hat{C}\hat{f}}\) locus, choosing \(\fcut\) at the value that corresponds to SNR \(\rhostar\), as defined by Equation~\eqref{eqn:fcut}, would include a triangular-shaped region of \((\hat{f},\hat{\sigma})\) space that does not exist in the catalogue, and hence the model would expect a set of stars that is not included in the data. This is the discrepancy discussed in Section~\ref{sec:Real data}. The flux cut, \(\fcut\), must be larger than the flux of the intersection between the fuzzy locus and the SNR cut line. In the 11 NGC~4258 fields, this forces \(\rhostar>5.4\). 

However, using a small cut value does not help the inference of \(\fTRGB\), which is uninformed by stars \(\gg \sigma\) away from it. The posterior is \((N+5)\)-dimensional, and hence increasing \(\fcut\) and reducing the number of latent fluxes to infer will increase the speed of the sampler. Furthermore, we expect to see more low-flux contamination of the RGB if we use a low \(\fcut\), and the RGB LF will eventually no longer follow the power law of Equation~\eqref{eqn:full LF}. 

We test the robustness of the inference of \(\fTRGB\) with varying \(\rhostar\). For simulated catalogues, using population-level parameter truths from Section~\ref{sec:coverage}, we find no bias for different choices of \(\rhostar\), provided the cut lies at least \(\sim\!5\sigma\) fainter than the true tip. For cuts higher than this, the tip becomes unidentifiable, and the HMC chains fail to converge. 

For the CATs NGC~4258 data, we fit across a range of cuts and find that we need at least 300 stars in the LF for each tip to be identifiable. This defines a per-field upper bound on the cuts we can choose. The lower bound for retained fits is given by \(\rhostar>5.4\). Further, some of the fits do not pass our convergence criteria of \(\hat{R} < 1.01\) and no divergences, and these fits are not retained. For each field, we find the \(\rhostar\) range where the inference of \(\fTRGB\) is stable. We define this stable region as the longest contiguous region where \(\mTRGB\) varies by less than the median uncertainty of the retained cuts. If two or more regions are found with the same width, we choose the one with the highest \(\rhostar\). The adopted \(\rhostar\) is the closest sampled value to the midpoint of the stable plateau, with ties resolved upward. The results are shown in Table~\ref{tab:rho_sweep}. 

\begin{table*}
\centering
\caption{Range of flux cuts over which each field-level TRGB magnitude is stable. Here, \(d_{\mathrm{gal}}\) is the distance to the centre of the galaxy \citep{scolnic_cats_2023}, deprojected into the disc plane, and \(\Delta \mTRGB\) is the full range spanned by the median \(\mTRGB\) across the sampled cuts, given separately for the stable and retained regions.}
\label{tab:rho_sweep}
\begin{tabular}{lcccccccc}
\toprule
 & & \multicolumn{3}{c}{\(\rhostar\)} & \(N\) & \multicolumn{2}{c}{\(\Delta \mTRGB\) (mag)} & Tolerance \\
\cmidrule(lr){3-5}
\cmidrule(lr){7-8}
Field & \(d_{\mathrm{gal}}\) (kpc) & Adopted & Stable & Retained & Range & Stable & Retained & (mag) \\
\midrule
NGC~4258-1    & 18.5 & 24   & 21--26     & 16--26   & 532--2764  & 0.026 & 0.084 & 0.027 \\
NGC~4258-2    & 26.3 & 14   & 12.5--16   & 10--16   & 366--625   & 0.017 & 0.439 & 0.186 \\
NGC~4258-3    & 34.8 & 8    & 6--10.5    & 6--10.5  & 313--508   & 0.034 & 0.034 & 0.141 \\
NGC~4258-4 G1 & 29.2 & 17   & 13--21     & 10--21   & 431--1460  & 0.091 & 0.445 & 0.113 \\
NGC~4258-4 G2 & 33.1 & 11   & 6--16.5    & 6--16.5  & 301--806   & 0.056 & 0.056 & 0.067 \\
NGC~4258-5    & 22.9 & 11   & 10--11.5   & 6--12    & 1055--3846 & 0.056 & 0.293 & 0.061 \\
NGC~4258-6    & 25.9 & 11   & 10.5--11.5 & 6--12    & 813--5117  & 0.029 & 0.338 & 0.045 \\
NGC~4258-7    & 21.6 & 27.5 & 22.5--30   & 10--33   & 300--4389  & 0.012 & 0.127 & 0.024 \\
NGC~4258-8    & 38.9 & 15   & 6--22.5    & 6--25    & 417--2223  & 0.017 & 0.037 & 0.026 \\
NGC~4258-9    & 44.3 & 12.5 & 7.5--15    & 7.5--15  & 326--842   & 0.009 & 0.009 & 0.033 \\
NGC~4258-10   & 33.2 & 15   & 7--25      & 6--25    & 449--2905  & 0.013 & 0.018 & 0.016 \\
\bottomrule
\end{tabular}
\end{table*}

We find particularly narrow stable regions for Fields~5 and 6. We do not impose a minimum width on the stable region, but if we were to include one, it could be argued that these fields do not display a sufficiently wide plateau. Therefore, we report \(\MTRGBHSTband\) values with and without these fields in Section~\ref{sec:Real data} for comparison with \cite{li_standardized_2023}. GO-16198 was ``designed to increase the sample of Cepheids in NGC~4258'' \citep{anand_comparing_2022}, and so images a region of higher surface density, and hence more crowding and blending of stellar point spread functions \citep{freedman_status_2025}. We find a correlation between the deprojected distance of a field from the centre of the galaxy and the drift of \(\mTRGB\) across the retained \(\rhostar\) region. The six innermost fields (1, 2, 4~G1, 5, 6, 7), at 18.5--29.2~kpc, drift by \(\ge 0.084\)~mag, whereas the five outermost fields, at 33.1--44.3~kpc, drift by \(\le 0.056\)~mag, with no overlap. The effect holds even in a single programme: the two innermost of the four fields of GO-10399, Fields~2 (26.3~kpc) and 4~G1 (29.2~kpc), drift by 0.439 and 0.445~mag, respectively, over the retained range, whereas the outermost, Fields~3 (34.8~kpc) and 4~G2 (33.1~kpc), drift by 0.034 and 0.056~mag, respectively. 

The largest drifts occur in fields whose \(\fTRGB\) posterior is multimodal in part of the retained range. For Fields~2 and 4~G1, single steps of 0.292 and 0.235~mag account for most of the drift, as posterior density transfers from a fainter mode to a brighter one.

We find that the drift is consistent with curvature in the RGB LF away from the single power law (Equation~\ref{eqn:full LF}) that the model expects. For the NGC~4258 fields, we find that the RGB LF slope increases with \(\rhostar\) over the retained range for all 11 fields, with a median increase in \(a\) of 0.84. The maximum increase is found in Field~6, where \(a\) is inferred to be 1.45 at \(\rhostar=6\) and 5.28 at \(\rhostar=12\), an increase of 3.83. The cut moves the tip position because the RGB LF slope, \(a\), is positively correlated with the tip, as seen in Figure~\ref{fig:corner_NGC4258-5}. For the 11 fields, the variation in \(a\) and in the inferred tip magnitude are correlated (Spearman rank correlation \(+0.71\), \(p=0.015\)). Restricting to the stable region reduces the median increase in \(a\) to 0.52 and removes the correlation (Spearman rank correlation \(+0.20\), \(p=0.56\)). We are able to produce drifts of up to 0.25~mag by adding curvature to the RGB LF in simulated datasets, consistent with all but the four fields with the largest drift. \cite{makarov_tip_2006} also find this dependence, noting that their fitting for the population-level parameters shows a coupling between the RGB slope and the inferred tip magnitude; a shallower slope prefers a fainter tip. They attribute this coupling to incompleteness affecting the fit when the tip lies within \(\sim\!1\)~mag of the photometric limit. However, the TRGB of each field analysed here lies 1.8--3.0~mag brighter than the \(\mathrm{SNR} = 4\) CATs limit. 

The origin of the curvature is not identified here. \cite{conn_bayesian_2011} consider a parametrisation where the RGB sits on top of a contaminant population that continues above the tip. This would lead to a decreasing RGB LF slope towards the tip, in the opposite sense to the curvature that we see in the NGC~4258 fields. Even without knowing the source of the curvature, we can reduce its effect by making a cut at a sufficiently high \(\rhostar\). This is what the plateau represents: a region where the curvature is stable and \(\fTRGB\) settles. 

We can also quantify its effect on our final TRGB magnitudes. We still observe a consistent drift to brighter magnitudes in the 11 fields over their stable regions: if we choose the lowest cut in the stable region for each field, we obtain a value of \(\mgalaxy=25.338 \pm 0.027\)~mag, and \(\tau=0.076^{+0.026}_{-0.019}\)~mag, whereas if we choose the highest cuts, we find \(\mgalaxy=25.314 \pm 0.026\)~mag, and \(\tau=0.072^{+0.024}_{-0.018}\)~mag. These two cuts represent the two extremes, resulting in a difference of 0.024~mag, which is comparable to the uncertainty in the galaxy-level magnitude. The scatter \(\tau\) is appreciably greater than zero for both cut choices, so the field-to-field scatter is not an artefact of the cut policy. 

Having found that the cut policy affects the final galaxy TRGB magnitude, we include an uncertainty of \(\pm 0.012\)~mag in our error budget (Table~\ref{tab:error_budget}), which corresponds to half of the maximum difference that the cut choice can make.

\end{document}